\documentclass[tradiabstract]{aa}   

\usepackage{graphicx}
\usepackage{txfonts}
\usepackage{epsfig}
\usepackage{longtable}
\usepackage{lscape}
\usepackage{amssymb}
\usepackage[colorlinks=true,citecolor=blue]{hyperref} 
\usepackage{refcount}
\usepackage{txfonts}
\bibpunct{(}{)}{;}{a}{}{,} 
\usepackage{xcolor}

\def\src {SDSS~J164100.10$+$345452.7}
\def\corto {J1641}
\def\MH{Mets\"ahovi}

\def \gnls {{$\gamma$-NLS1}}

\newcommand\kms{\ifmmode {\rm km\ s}^{-1} \else km s$^{-1}$\fi} 
\newcommand\FWHM{\ifmmode {\rm FWHM} \else ${\rm FWHM}$\fi}
\newcommand\Lsun{\ifmmode L_{\odot} \else $L_{\odot}$\fi} 
\newcommand\Msun{\ifmmode M_{\odot} \else $M_{\odot}$\fi} 
\newcommand\Hbeta{\ifmmode {\rm H}\beta 
 \else H$\beta$\fi}

\def \ergsec{\hbox{erg s$^{-1}$}}

\def \hcm {\hbox {\ifmmode $ atom cm$^{-2}\else atom cm$^{-2}$\fi}}
\def \arcmin {\hbox{$^\prime$}}
\def \arcsec {\hbox{$^{\prime\prime}$}}

\def \gray {$\gamma$-ray}

\def \sw {{\em Swift}}

\def \fermi{{\it Fermi}}

\def \apj {ApJ}
\def \apjl {ApJL}
\def \apjs {ApJS}
\def \aap {A\&A}

\def \mnras {MNRAS}

\def \ssr {SSRv}

\begin{document} 
\title{Long-term {\em Swift} and \MH{} monitoring of  SDSS~J164100.10$+$345452.7 reveals multi-wavelength correlated variability}
\titlerunning{Long-term monitoring of  SDSS~J164100.10$+$345452.7}
\authorrunning{Romano et al.}
%
%
  \author{
  P.~Romano\inst{\ref{oab}}
   \and A.~L\"ahteenm\"aki\inst{\ref{amh},\ref{auni}}  
   \and S.~Vercellone\inst{\ref{oab}}
    \and L.~Foschini\inst{\ref{oab}}
    \and M.~Berton\inst{\ref{eso}}
    \and C.M.~Raiteri\inst{\ref{oat}}
   \and V.~Braito\inst{\ref{oab},\ref{cathol},\ref{trento}}
   \and S.~Ciroi\inst{\ref{unipd}}
   \and E.~J\"arvel\"a~\inst{\ref{esa},\ref{uoo}} 
    \and S.~Baitieri\inst{\ref{unibi},\ref{oab}}  
   \and  I.~Varglund\inst{\ref{amh},\ref{auni}} 
    \and M.~Tornikoski\inst{\ref{amh}}
    \and S.~Suutarinen\inst{\ref{amh}} 
 }
  
 \institute{
   INAF, Osservatorio Astronomico di Brera, Via Emilio Bianchi 46, I-23807 Merate (LC), Italy\label{oab}  
   \email{patrizia.romano@inaf.it}\label{mainlab} 
   \and Aalto University Mets\"ahovi Radio Observatory, Mets\"ahovintie 114, 02540 Kylm\"al\"a, Finland\label{amh} 
   \and Aalto University Department of Electronics and Nanoengineering, PO Box 15500, 00076 Aalto, Finland\label{auni}
   \and European Southern Observatory (ESO), Alonso de C\'ordova 3107, Casilla 19, Santiago 19001, Chile \label{eso}
  \and INAF, Osservatorio Astrofisico di Torino, Via Osservatorio 20, 10025 Pino Torinese (TO), Italy\label{oat}
 \and Department of Physics, Institute for Astrophysics and Computational Sciences, The Catholic University of America, Washington, DC 20064, USA\label{cathol}
 \and Dipartimento di Fisica, Universit\`a di Trento, Via Sommarive 14, Trento 38123, Italy\label{trento}
   \and Dipartimento di Fisica e Astronomia, Universit\`a di Padova, 35122 Padova, Italy\label{unipd}
   \and European Space Agency, European Space Astronomy Centre, C/ Bajo el Castillo s/n, 28692 
            Villanueva de la Cañada, Madrid, Spain \label{esa}
   \and Homer L. Dodge Department of Physics and Astronomy, The University of Oklahoma, 440 W. Brooks St., Norman, OK 73019, USA\label{uoo}
   \and Dipartimento di Fisica G.\ Occhialini, Universit\`a degli Studi di Milano Bicocca, I-20126 Milano, Italy\label{unibi}
}

  \date{Received 18 January 2023; accepted 08 March 2023}

\abstract
{We report on the first multi-wavelength {\em Swift} monitoring campaign performed on SDSS~J164100.10$+$345452.7, 
a nearby narrow-line Seyfert 1 galaxy formerly known as radio quiet which was however recently detected both in the radio (at 37\,GHz)  
and in the $\gamma$-rays, a behaviour which hints at the presence of a relativistic jet. 
During our 20-month {\em Swift} campaign, 
while pursuing the primary goal of assessing the baseline optical/UV and X-ray properties of SDSS~J164100.10$+$345452.7, 
we caught two radio flaring episodes, one each year. 
Our strictly simultaneous multi-wavelength data closely match the radio flare and allow us to 
unambiguously link the jetted radio emission of SDSS~J164100.10$+$345452.7. 
Indeed, for the X-ray spectra preceding and following the radio flare 
a simple absorbed power-law model is not an adequate description, and an extra absorption 
component is required. 
The average spectrum of SDSS~J164100.10$+$345452.7 can be best  described 
by an absorbed power law model with a photon index $\Gamma=1.93\pm0.12$, 
modified by a partially covering neutral absorber with a covering fraction $f=0.91_{-0.03}^{+0.02}$. 
On the contrary, the X-ray spectrum closest to the radio flare does not require such extra absorber and 
is much harder ($\Gamma_{\rm flare} \sim 0.7\pm0.4$), thus 
implying the emergence of a further, harder spectral component. 
We interpret this as the jet emission emerging from a gap in the absorber.   
The fractional variability we derive in the optical/UV and X-ray bands are 
found to be lower than the typical values reported in the literature, 
since our observations of \src{} are dominated by the source being in a low state, 
as opposed to the literature, where the observations were generally taken as a 
follow-up of bright flares in other energy bands. 
Under the assumption that the origin of the 37\,GHz radio flare is the emergence of a jet from an obscuring screen
also observed in the X-rays, the derived total jet power is $P^{\rm tot}_{\rm jet}=3.5\times10^{42}$\,erg\,s$^{-1}$, 
comparable to the lowest measured in the literature. \\
} 


   \maketitle
%

  \section{Introduction \label{sdss1641:intro}}

%
Narrow-Line Seyfert 1 galaxies (NLS1s) are active galactic nuclei (AGN) characterised in the optical regime 
by narrow permitted emission lines (\Hbeta{} \FWHM $<$ 2000 $\kms$, \citealt[][]{Goodrich1989:nls1def}),  
weak forbidden oxygen lines (flux ratio [\ion{O}{III}]\,$\lambda 5007/ \Hbeta < 3$), 
and often strong iron emission lines (high \ion{Fe}{II}/\Hbeta, \citealt[][]{OsterbrockP1985:nls1def}),
properties that distinguish them from the population of Seyfert 1 galaxies (broad-line Seyfert 1s, BLS1s). 
Indeed, NLS1s are part of the so-called population A of the AGN main sequence 
\citep[][]{Sulentic2002,SulenticMarziani2015:FrASS,Marziani2018:FrASS}. 
 In the soft X-rays NLS1s also have extreme properties, i.e., steep spectra 
 (\citealt[][]{Brandt1997a};  \citealt[][]{Leighly1999b}, $\Gamma_{\rm NLS1}= 2.19\pm0.10$ vs $\Gamma_{\rm BLS1}= 1.78\pm0.11$)  
 fast and large-amplitude variability 
\citep[][]{BollerBF1996:softX}, with some showing X-ray flares up to a factor of 100 in flux, 
on time-scales of days, while BLS1s are generally seen to vary by a factor of a few.  
These distinctive properties \citep[e.g.][]{Peterson2004} are customarily understood in terms of 
low  mass ($10^6$--$10^8$ M$_{\sun}$)  
central black holes and higher accretion rates, close to the Eddington limit 
(but see, \citealt{Viswanath2019:NLS1masses}), hence possibly in an early stage of evolution \citep[][]{Mathur2000}. 
A small fraction of NLS1s (4--7\,\%, \citealt[][]{Komossa2006:rlnl1q,Cracco2016})  
have been found to be 
radio loud\footnote{We remark that the radio loudness parameter as defined above is becoming more and 
more inaccurate and unreliable in properly describing these sources, as recently summarised by \citet[][]{Berton2021:AN}.} 
($S_{\rm radio}/S_{\rm optical} > 10$) 
and show a flat radio spectrum 
(\citealt[][]{Oshlack2001:pks2004-447,Zhou2003:0948,Yuan2008}; 
see also, \citealt[][]{Lahteenmaki2017}). 
Additionally, some show a hard spectral component 
and hard X-ray spectral variability \citep[][]{Foschini2009:Adv}.

The further, smaller subclass of $\gamma$-ray emitting NLS1 (\gnls) galaxies 
was recently added to the picture, as a consequence of their detection 
at high energies (E $> 100$\,MeV) by {\it Fermi}-LAT 
\citep[PMN~J0948$+$0022,][]{Abdo2009:J0948discov,Foschini2010:J0948,Abdo2009:J0949mw}. 
The \gnls{} now include about 35 objects \citep[e.g.][]{Romano2018:nls1_cta,Foschini2021:new_sample,Foschini2022:new_sample2} 
whose properties strongly resemble those of jetted sources  
\citep[see, e.g.][]{Foschini2012:review,Foschini2015:fsrl_nls1,Dammando2016:jets_nls1}.
The  calculated jet power \citep[$10^{42.6}$--$10^{45.6}$\,erg\,s$^{-1}$,][]{Foschini2015:fsrl_nls1}, 
for instance,  is lower than that of flat spectrum radio quasars (FSRQs) 
and partially overlapping those of BL Lacertae objects (BL Lac). 
Furthermore, the spectral energy distributions (SEDs) of $\gamma$-NLS1 galaxies resemble 
the typical double-humped one of those of other jetted sources such as FSRQs 
and BL Lacs, as reported in~\citet[][]{Abdo2009:J0948discov} and \citet[][]{Abdo2009:J0949mw} 
and more recently in~\citet[][]{Foschini2010:J0948} and \citet[][]{Foschini2015:fsrl_nls1}. 
In particular, Figure~1 in~\cite{Foschini2010:J0948} shows the SED of PMN~J0948$+$0022 
as compared to the average SEDs of FSRQs and BL Lacs and of those of representative radio galaxies 
such as NGC~6251, M~87, and Centaurus A. 
The SED of the prototypical $\gamma$-NLS1 PMN~J0948$+$0022 lies in-between those of 
high-power FSRQs and low-power BL Lacs, not surprisingly according to the value of the jet 
power reported above. 
Since the radio luminosity function of \gnls{} is the continuation at low luminosity of that of FSRQs, 
it is possible that \gnls{} are FSRQs in an early stage of their life cycle \citep[][]{Berton2016c,Berton2017:3C286}         
and the possible differences in observed host galaxies 
could be explained by ongoing or recent mergers  (\citealt[e.g.][]{Jarvela2018,Paliya2020,Shao2022}, but see \citealt[][]{Varglund2022} 
where only one jetted NLS1 shows signs of interaction).

\object{SDSS~J164100.10$+$345452.7} \citep[][]{Zhou2006:SDSS_NLS1s}, hereon \corto, 
is a nearby NLS1 ($z=0.16409\pm0.00002$, \citealt[][]{Albareti2017:SDSS13DR}) 
hosted in a spiral galaxy \citep{Olguin-Iglesias2020_HostClass}, 
initially classified as radio quiet.  
\citet[][]{Lahteenmaki2018}, who observed this source 71 times at 37\,GHz 
with the 13.7\,m radio telescope at Aalto University \MH{}  Radio Observatory, 
however reported a detection of \corto{} twice (detection percentage of 2.8\,\%), 
with an average flux of $S_{\rm 37\,GHz, ave} = 0.37$\,Jy, and a maximum flux of 0.46\,Jy.  
Since such detections suggested the presence of jets, they also sought \corto{} 
in the {\it Fermi}-LAT data and obtained a 
$S_{\rm E>100\,MeV} = (12.5\pm2.18)\times10^{-9}$\,ph\,cm$^{-2}$\,s$^{-1}$ 
(with a maximum-likelihood test statistics TS$=39$), making this a radio emitting \gnls. 
This detection, combined with the \MH{}  detection of six more radio-silent NLS1s and one more radio-quiet NLS1, 
puts a severe strain on the belief that so-called radio-quiet 
or radio-silent NLS1s hosted in spiral galaxies are unable to launch jets. 
New radio observations at low frequencies suggest that the emission of the jet in these objects 
is absorbed, possibly via a free-free mechanism \citep[][]{Berton2020:absorbed_jets,Jarvela2021:FrASS}. 
Therefore, NLS1s may harbour a previously unknown class of relativistic jets, completely undetectable at low radio frequencies, 
that is where most observations have been carried out to date.
\corto{} is now part of the sample of approximately one hundred NLS1s frequently monitored at 37\,GHz at \MH{} 
\citep[][]{Lahteenmaki2017}.  

In this paper we report on a 2-year multi-wavelength (optical, ultra-violet, and X-ray) 
monitoring of \corto{} with the Neil Gehrels {\it Swift} Observatory \citep[][]{Gehrels2004} 
that was performed in  2019--2021 simultaneously with 
37\,GHz observations obtained as part of the \MH{} NLS1 monitoring program. 
In particular, we focus on a 37\,GHz flare observed on 2020-05-24 to 2020-05-26. 
In Sect.~\ref{sdss1641:data} we describe the observations and  the data reduction; 
in Sect.~\ref{sdss1641:results} we describe our analysis and present our results and 
in Sect.~\ref{sdss1641:discussion} we discuss their implications.  
We adopt the usual $\Lambda$-cold dark matter ($\Lambda$-CDM) cosmology~\citep{Komatsu2011_cosmology} with H$_{\rm 0}$ = 70\,${\rm km}\, {\rm s}^{-1}\,{\rm Mpc^{-1}}$, $\Omega_{m}$ = 0.27, and $\rm \Omega_{\Lambda}= 0.73$ to facilitate direct comparison with the results in \cite{Foschini2015:fsrl_nls1}.

   \section{Observations and data reduction \label{sdss1641:data}}

\begin{figure*} 
\vspace{-0.7truecm}

\hspace{-0.3truecm}
 \includegraphics[angle=270,width=1.11\textwidth]{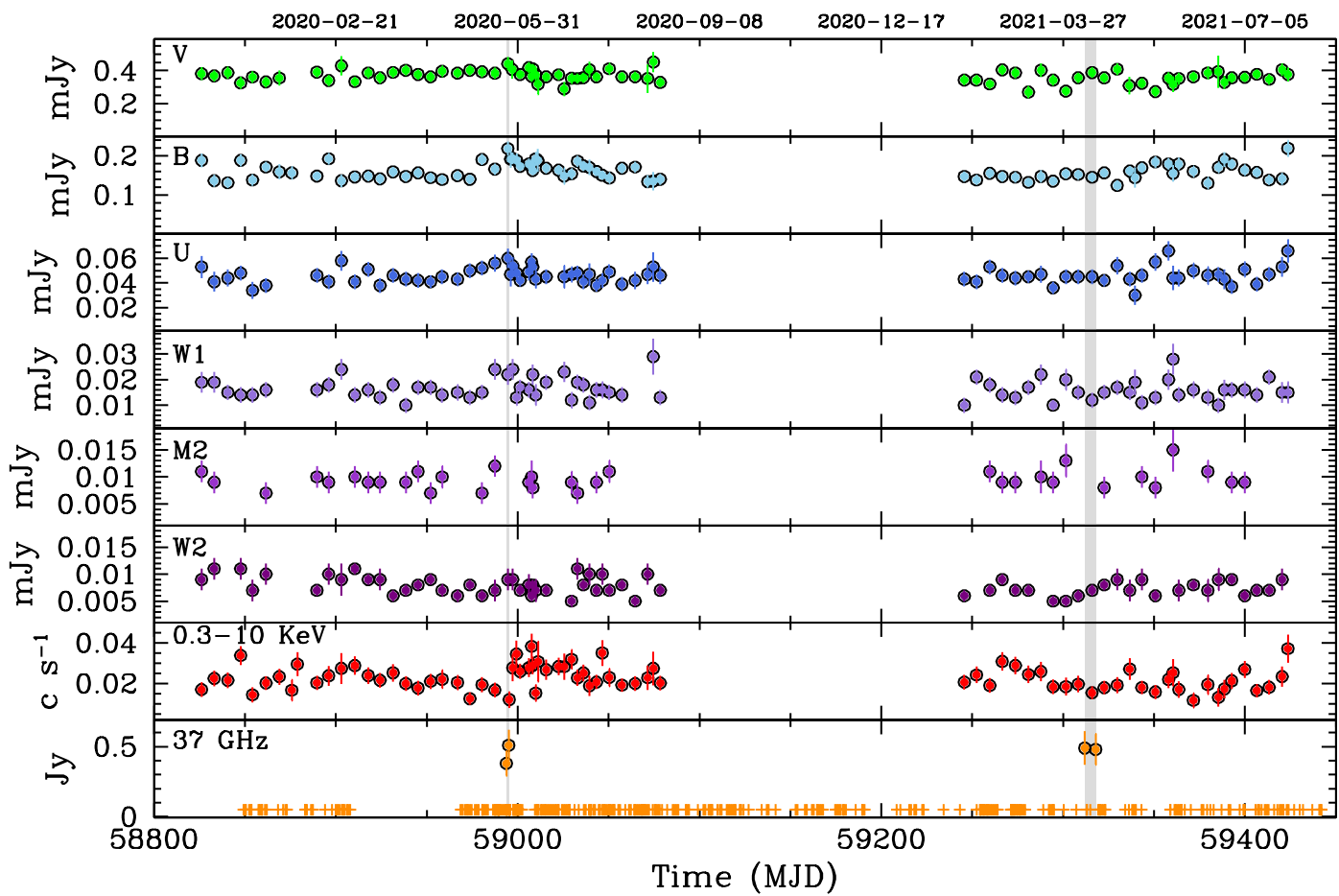}
\vspace{-2.3truecm}
  \caption{\label{sdss1641:fig:multi}
Multi-wavelength light curves of \src.  
The optical, UV, and X-ray light curves were collected by \sw{} 
from 2019-12-09 to 2020-08-17 (first year campaign), 
from 2021-01-31 to 2021-07-28 (second year), and are shown with 1\,$\sigma$ errorbars. 
The data at 37\,GHz were collected at Mets\"ahovi  (<4\,$\sigma$ non-detections represented by crosses).   
The grey bands mark the Mets\"ahovi detections.
The top axis reports representative dates during the campaigns.
} 
\end{figure*} 

                 \subsection{Swift \label{sdss1641:data_swift}}  

The \sw\ data were collected through two yearly monitoring campaigns (Target ID 11395, PI: P.\ Romano) 
with a pace of one $\sim2$--3\,ks observation per week 
from 2019-12-09 to 2020-08-17 (97\,ks) 
and from 2021-01-31 to 2021-07-28 (68\,ks) with the X--ray Telescope \citep[XRT,][]{Burrows2005:XRT} and 
the UV/Optical Telescope \citep[UVOT,][with the ``U+B+V+all U'' filter setup]{Roming2005:UVOT}, 
for a total exposure of 165\,ks (68 observations).  
The pace is ideal for variability studies, since the data collected are a 
regular and casual sampling of the light curve of the source at a 
resolution $\sim1$\,week, irrespective of the source flux state.
We note that each observation may consist of 1--3 snapshots (satellite orbits) 
which are just a few hours apart. 

Additionally, three target of opportunity (ToO) observation campaigns 
(ObsIDs: 00011395030, -032, -033; 00011395035, -036; 00011395038, -039, -042, -044, -046, -048) 
were obtained (PI: P.\ Romano) in response to a radio detection from \MH{} on 2020-05-24 
(MJD 58993--58995, see below Sect.~\ref{sdss1641:data_mh}), for a total of 16\,ks (11 observations). 
We note that the observing pace of these ToOs is much denser than that of the monitoring campaigns,
so when timing is performed we distinguish the case of  the campaign-only dataset from the full dataset.  
No ToO observations were obtained after the second set of radio detections 
recorded in 2021. 
Tables~\ref{sdss1641:tab:swift_xrt_log} and \ref{sdss1641:tab:swift_xrt_log2} report
the log of the \sw/XRT observations for the first and second year, respectively, 
and include the observing sequence (ObsID), date (MJD of the middle of the observation), 
start and end times (UTC) of each observation, and XRT exposure time.

The \sw/XRT data were reduced and analysed by using standard procedures within 
{\sc FTOOLS}\footnote{\href{https://heasarc.gsfc.nasa.gov/ftools/ftools_menu.html}
{https://heasarc.gsfc.nasa.gov/ftools/ftools\_menu.html.}} (v6.29b),  
and the  calibration database 
CALDB\footnote{\href{https://heasarc.gsfc.nasa.gov/docs/heasarc/caldb/caldb_intro.html}
{https://heasarc.gsfc.nasa.gov/docs/heasarc/caldb/caldb\_intro.html.}}  (v20211105).  
In particular, the XRT data were processed and filtered with the task {\sc xrtpipeline} (v0.13.6). 
The XRT light curve was generated in the 0.3--10\,keV energy range with one point per observation
and one per snapshot.
We used {\sc sosta} within {\sc XIMAGE} and adopted the background measured in an annular region with 
an inner radius of 80 pixels and an external radius of 120 pixels centred at the source position. 
The PSF losses and vignetting were corrected for by using the appropriate exposure maps. 
In case of a non detection, a 3\,$\sigma$ upper limit was calculated by using 
{\sc sosta} and {\sc uplimit} within {\sc XIMAGE} and the background defined above. 
The XRT 0.3--10\,keV light curve, binned at one bin per observation, 
(which corresponds to one point per week during the monitoring campaign 
and several points per week during the ToOs)  is shown in Fig.~\ref{sdss1641:fig:multi}.  
We note that close to the 2020 radio flare \corto{} was observed 
on 2020-05-25 (MJD 58994, observation 029 for 1.7\,ks) 
and on 2020-05-27 (MJD 58996, observation 030 for $\sim 350$\,s).  
Although individually, these observations did not yield a detection
($\sim1.1\times10^{-2}$ counts\,s$^{-1}$ at the 2.8\,$\sigma$ level,  
and $\sim 0.01$ counts\,s$^{-1}$ at the 1.4\,$\sigma$ level, respectively), 
their combination did, at $(1.2\pm0.4)\times10^{-2}$ counts\,s$^{-1}$, so we 
used this combination as opposed to two upper limits.  
On the contrary, we discarded observation 069 due to its low exposure ($\sim 350$\,s) 
and no close observations to combine it with. 
For the spectroscopic analysis of the XRT data, source events were extracted from a circular region with 
a 20 pixel radius; ancillary response files were generated with  the task {\sc xrtmkarf} 
to account for different extraction regions, vignetting, and PSF corrections; 
the spectral redistribution matrices in CALDB were used. 
The analysis of the UVOT data, which were collected in all filters (optical $v$, $b$, $u$ and UV $w1$, $m2$, $w2$),  
was performed with the  tasks  {\sc uvotmaghist},   {\sc uvotimsum}, and {\sc uvotsource} included in {\sc FTOOLS}
(v6.31 and  {\sc CALDB} v20211108, which offer a check for frames affected by small scale sensitivity, SSS, 
issues\footnote{\href{https://swift.gsfc.nasa.gov/analysis/uvot_digest/sss_check.html}{https://swift.gsfc.nasa.gov/analysis/uvot\_digest/sss\_check.html}.}). 
{\sc uvotmaghist} was used to identify the frames of each UVOT image  affected by SSS issues 
and to generate a light curve in each filter at the frame level (the best time resolution available), 
{\sc uvotimsum} to sum the sky images for {\sc uvotsource} to calculate the magnitude of the source 
through aperture photometry within a circular region centred on the best source position 
by applying  the required corrections related to the specific detector characteristics. 
We adopted circular regions with radii of 5\,\arcsec{} and 18\,\arcsec{}  for the source and background,
respectively. We also discarded the source detections that yielded magnitudes with an error $>0.3$\,mag. 
The observed light curves (in units of mJy) binned at one point per observation
are shown in the six top panels of Fig.~\ref{sdss1641:fig:multi}.

                 \subsection{\MH \label{sdss1641:data_mh}}  
The Aalto University \MH{} Radio Observatory in Finland operates  a 13.7\,m radio telescope, 
used for monitoring of AGN at 22 and 37\,GHz. 
The NLS1 monitoring program, fully described in \citet{Lahteenmaki2017}, started in 2012 and is currently ongoing. 
The measurements are carried out with a 1\,GHz band dual-beam receiver centred at 36.8\,GHz, 
with typical integration times between 1200 and 1800\,s. 
The detection limit is on the order of 0.2\,Jy under optimal conditions.
Data points with a signal-to-noise ratio < 4 are handled as non-detections.
They may occur either because the source is too faint or because of compromised 
weather conditions. In the latter case the measurement is discarded.  
Fainter sources, such as NLS1s, commonly flicker around the detection limit, 
causing frequent non-detections among the possibly rare detections that 
occur only when the peak of a flare is seen.
The main flux calibrator is DR21, with NGC~7027, 3C~274, and 3C~84 used as secondary calibrators. 
The full data reduction procedure is reported in \citet{Teraesranta98}.

\corto{} is normally observed at least once a week and during more intensive periods, for example, 
multi-frequency campaigns or flares, even daily. It has now been detected several times 
since the early reports in \citet[][]{Lahteenmaki2018}, 
with a maximum flux density of 0.65$\pm$0.12\,Jy on 2019-09-07.  
In Fig.~\ref{sdss1641:fig:multi}, bottom panel,  we show the data simultaneous with the \sw{}  campaigns. %
Four detections were achieved,  on 
2020-05-24 (MJD 58993.7820, flux density of 0.38$\pm$0.09\,Jy),  
2020-05-26 (MJD 58995.0987, 0.51$\pm$0.11\,Jy), 
2021-04-07 (MD 59311.9087, 0.49$\pm$0.12\,Jy), and 
2021-04-13 (MJD 59317.8955, 0.48$\pm$0.11\,Jy). 
We shall hereon refer to the May  2020 detections as the 2020 flare, and 
the April 2021 ones as the 2021 flare. 
Further detections were obtained after the end of the \sw{} campaigns on 
2021-08-31 (MJD 59457.7556, 0.44$\pm$0.09\,Jy) and 
2021-09-09 (59466.7323, 0.48$\pm$ 0.09\,Jy).

  \section{Data analysis and results \label{sdss1641:results}}

           \subsection{Optical, UV, and X-ray variability}\label{sdss1641:results_variab} 
\setcounter{table}{0}
\begin{table}
\tabcolsep 4pt
 \begin{center}
   \caption{Fractional variability in different energy bands. 
\label{sdss1641:tab:Fvar}}
    \begin{tabular}{lll ll }
  \hline
 \hline
 \noalign{\smallskip}
                              & \multicolumn{2}{c}{ObsID} & \multicolumn{2}{c}{Snapshot}  \\
  \noalign{\smallskip}
        Filter             & F$_{\rm var}^{\rm camp}$ & F$_{\rm var}^{\rm full}$ &  F$_{\rm var}^{\rm camp}$ & F$_{\rm var}^{\rm full}$ \\
  \noalign{\smallskip}
 \hline
 \noalign{\smallskip}
          V            	& -- & --                                              & -- & -- \\
          B            	& $0.09 \pm 0.01$ & $0.09\pm0.01$  & $0.10\pm0.02$ & $0.11\pm0.01$ \\
          U            	& $0.04 \pm 0.06$ & --                       & $0.02\pm0.11$  &--   \\
          W1           	& $0.13 \pm 0.04$ & $0.13\pm0.04$   & -- & --   \\
          M2           	& -- & --                                              & -- & -- \\
          W2           	& $0.07 \pm 0.05$ & $0.07 \pm 0.05$ & $0.11\pm0.06$  & $0.10\pm0.06$ \\
          X-ray      	& $0.16 \pm 0.03$ & $0.16 \pm 0.03$ & $0.11\pm0.04$ & $0.13\pm0.04$\\ 
  \noalign{\smallskip}
 \hline
 \noalign{\smallskip}
    \end{tabular}
 \end{center}
  {\bf Notes.} The X-ray band is 0.3--10 keV.  
The values are calculated for light curves drawn from
campaign-only observations (F$_{\rm var}^{\rm camp}$) and the full set of observations (F$_{\rm var}^{\rm full}$),
binned both at the observation level  and at the snapshot level. 
Missing values are cases where variability is not well detected. 
 \end{table}

As shown in Fig.~\ref{sdss1641:fig:multi}, the \sw{} light curves show a degree of variability. 
To quantify it, we computed the fractional variability according to 
Eqs.~(10) and (B2)  in~\cite{Vaughan2003}, 
\begin{equation}
F_{\rm var} = \sqrt{\frac{S^{2} - \bar{\sigma^{2}}}{\bar{x}^{2}} } \pm 
\sqrt{\left(\sqrt{\frac{1}{2N}} \frac{\bar{\sigma}^{2}_{\rm err}}{\bar{x}^{2}F_{\rm var}}\right)^{2} +
\left(\sqrt{\frac{\bar{\sigma}^{2}_{\rm err}}{N}} \frac{1}{\bar{x}}\right)^{2}}.
\end{equation}
Since the ToO data (after MJD 58996) may introduce a bias towards the high/flaring states,   
we performed the calculation both over the full dataset ($F_{\rm var}^{\rm full}$, ToOs included) 
and by restricting to the campaign data only ($F_{\rm var}^{\rm camp}$).
Table~\ref{sdss1641:tab:Fvar} shows the results for both, and, since they are consistent 
within errors, from now on we shall consider the entire dataset for our analysis.   
We calculated $F_{\rm var}$ from the light curves binned both at the observation level and at the snapshot level. 
Missing values are due to cases where variability is not well detected \citep[see Appendix B in][]{Vaughan2003}. 
We note that the fractional variability measured in \corto{} is lower than the typical values reported 
in \citet[][]{Dammando2020:swift} for six well-known \gray{} NLS1s. 
This is unsurprising, since we focused on a long-term campaign, thus reducing the impact 
of high flux levels as a consequence of observations triggered by flares observed 
in other energy bands. 

 \setcounter{table}{1}
\begin{table}
\caption{Doubling and halving times. 
\label{sdss1641:tab:doubling}}    
 \tabcolsep 2pt
 \begin{center}
\small              
    \begin{tabular}{lll ll }
  \hline
 \hline
 \noalign{\smallskip}
                              & \multicolumn{2}{c}{ObsID} & \multicolumn{2}{c}{Snapshot}  \\
   \noalign{\smallskip}
      D/R                      &   R                             & D                           & R                           &        D        \\
  \noalign{\smallskip}
 \hline
 \noalign{\smallskip}
  $\tau_{\rm d}$         & 1.25                       & 1.70                        & 0.053                      & 5.90 \\ 
$\sigma(\tau_{d})$ & 3.9                         & 3.2                           &  5.0                        & 4.7 \\ 
$CR(t_{1})$             & $0.015\pm0.004$ & $0.029\pm0.004$  &  $0.012\pm0.003$ & $0.017\pm0.006$ \\
$CR(t_{2})$             & $0.031\pm0.010$ & $0.015\pm0.004$  &  $0.027\pm0.006$ & $0.045\pm0.013$\\
$t_{1}$                   & 59009.9084          &  59008.3476            &  59259.6634          & 59371.5824  \\ 
$t_{2}$                   &  59011.1743         & 59009.9084              &  59259.7270          & 59379.6139 \\
  \noalign{\smallskip}
 \hline
 \noalign{\smallskip}
\end{tabular}
 \end{center}
{\bf Notes.} The minimum doubling time (rise, R) and halving time (decay, D) $\tau_{d}$ are in units of days 
and their significance is $\sigma(\tau_{d})$. 
$t_{1}$ and $t_{2}$ are the times (MJD) when we measure the count rates $CR(t_{1})$  and 
$CR(t_{2})$ (count\,s$^{-1}$).   The values are calculated for light curves drawn from 
the data binned both at the observation level  and at the snapshot level.  
\end{table}

The minimum variability time scale in the X-ray energy band is 
$t_{\rm var} = {\rm ln}(\rm2) \times \tau_{\rm d}$\,days, 
where  $\tau_{\rm d}$ is the doubling/halving time defined by,  
\begin{equation}\label{EQ:tau_d}
F(t_{2}) = F(t_{1}) \times 2^{(t_{2} - t_{1})/\tau_{\rm d}},
\end{equation}
and $F(t_{1})$ and $F(t_{2})$ are the count rates at the times $t_{1}$ and $t_{2}$, respectively.
Given that the significance of $\tau_{d}$ is $\sigma(\tau_{d}) = |F(t_{1}) - F(t_{2})|/\sigma(F(t_{1}))$,
we selected doubling/halving times with $\sigma(\tau_{d}) \ge 3$. 
Table~\ref{sdss1641:tab:doubling} shows the minimum doubling/halving times 
and their significance when the light curves are binned both at the observation level and at the snapshot level. 
We also note that no difference was found when considering the full dataset or 
restricting to the campaign data only independently on the binning. 
Assuming  $t_{\rm var} = {\rm ln}(\rm2)\times(\min\{\tau_{\rm d}(R);\tau_{\rm d}(D) \})$, 
we obtain that the minimum variability timescale 
is $t^{\rm obs}_{\rm var} = 0.87$\,days and $t^{\rm snap}_{\rm var} =0.04$\,d  or $\simeq 1$\,hr
when the light curves are binned at the observation level and at the snapshot level,
respectively. 
n the comoving frame, $t\arcmin{}=t\,(1+z)$, 
we obtain $t\arcmin{}^{\rm obs}_{\rm var} = 0.75$\,days and $t\arcmin{}^{\rm snap}_{\rm var} =0.034$\,d, 
respectively.

The 0.3--10\,keV light curve shown in Fig.~\ref{sdss1641:fig:multi} is characterised by a dynamic range 
(maximum count rate / minimum count rate) of 3.3. 
To assess whether this time variability corresponds to detectable spectral variability throughout our monitoring, 
we  calculated the hardness ratio HR based on the 0.3--2\,keV and 2--10\,keV 
energy bands at the same time binning as the full light curve, and used it to test 
for spectral variability on timescales of $\sim 1$\,week (the pace of the campaign).  
We first fit the HR with a constant and obtained (2--10\,keV)/(0.3--2\,keV)$=0.48\pm0.02$ 
($\chi^2=83.4$, degrees of freedom d.o.f.$=71$, null-hypothesis probability nhp$=0.149$).
We then performed a linear fit, clearly consistent with the constant fit, 
since the first order coefficient is consistent with 0
($\chi^2=82.9$, d.o.f.$=70$, nhp=0.138). 
The F-test probability with respect to the constant is indeed 0.547 ($0.6\,\sigma$). 
The same procedure was applied to the whole set of observations available on \corto, thus including 
two archival observations performed on 2019-05-25 and  2019-08-29, 
and 9 observations we obtained in March 2022 as ToOs.
We obtained an F-test probability with respect to the constant of 0.660  ($0.95\,\sigma$) so we can conclude that 
we do not detect significant spectral variability on  timescales of $\sim 1$\,week. 
We note that the same conclusions are reached when the light curve is binned at the snapshot level
(few snapshots within a day, separated by about a week) with F-test probabilities below $0.5\,\sigma$.

              \subsection{X-ray time-selected spectroscopy\label{sdss1641:results_xrt}} 

\begin{figure} 
\vspace{+0.1truecm}

\hspace{-0.8truecm}
\includegraphics[angle=0,width=0.52\textwidth]{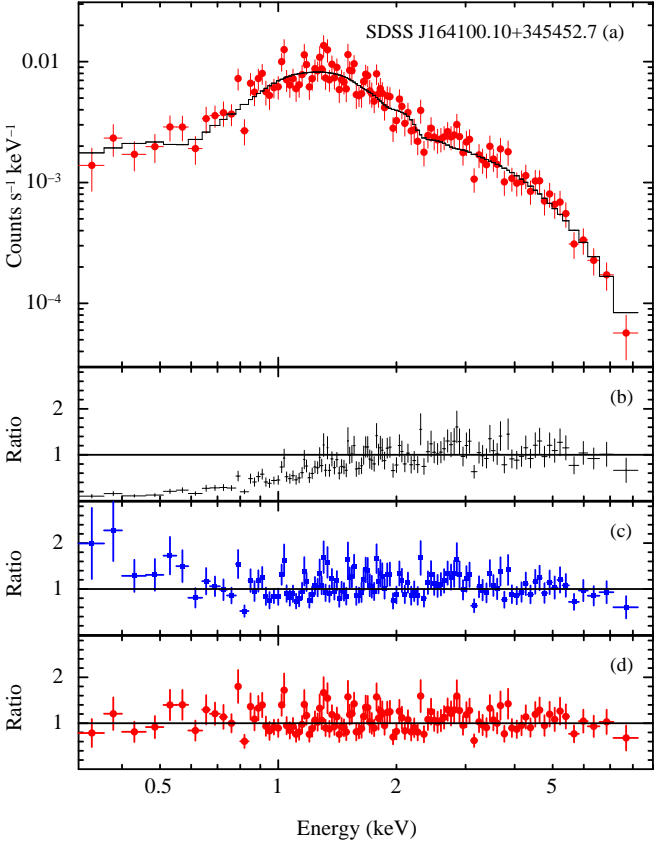}
\vspace{-0.5truecm}
  \caption{{\it Swift}/XRT average spectrum of \src. The data are drawn from the whole 2\,yr observing campaign 
 (details on the spectral fits can be found in Table~\ref{sdss1641:tab:swift_xrt_spec}). 
Panel (a) best fit obtained by adopting the model {\sc tbabs * zpcfabs * zpowerlw}; 
panel (b): data/model ratio from the fit with {\sc tbabs * zpowerlw} in the 2--10\,keV band; 
panel (c) data/model ratio from the fit with {\sc tbabs * ztbabs * zpowerlw} (0.3--10\,keV);
panel (d) data/model ratio from the fit with  {\sc tbabs * zpcfabs * zpowerlw} (0.3--10\,keV). 
\label{sdss1641:fig:tot_spec} 
}
 \end{figure} 

\setcounter{table}{2}
\begin{table}[t]
 \tabcolsep 4pt 
 \begin{center}
 \caption{Time-selected {\it Swift}/XRT spectroscopy. 
\label{sdss1641:tab:swift_xrt_spec}
}
 \begin{tabular}{l c c l } 
 \hline 
 \hline 
 \noalign{\smallskip} 
  Spectrum                                             &   {\bf total}                        & {\bf flare}    \\ 
\noalign{\smallskip} 
 \hline 
 \noalign{\smallskip}            
  Parameter &  Value  & Value	 & Units	    \\
  \hline 
 \noalign{\smallskip} 
 \multicolumn{4}{c}{{\tt tbabs * zpowerlw} (2--10\,keV)}        \\ 
 \noalign{\smallskip} 
$\Gamma_{\rm 2-10}$             & $1.84\pm0.15$       &  $0.94\pm1.35$  &    \\        
$F_{\rm 2-10}$                         & 6.24                         & 14.1                            &   $\times10^{-13}$ erg\,cm$^{-2}$\,s$^{-1}$   \\
$\chi^2$/d.o.f.                     & 42.40/48                  &  12.72/20 (70.48)       &   \\ 
nhp                                       &  0.701                      &   0.808                        &   \\ 
\noalign{\smallskip}
\hline 
 \hline 
 \noalign{\smallskip} 
\multicolumn{4}{c}{{\tt tbabs * zpowerlw} (0.3--10\,keV)}        \\ 
 \noalign{\smallskip} 
$\Gamma_{\rm 0.3-10}$           & $1.13\pm0.04$   &  $0.67_{-0.39}^{+0.38} $  &                \\
$F_{\rm 0.3-10}$                       & 10.3                            &  17.3                           &    $\times10^{-13}$ erg\,cm$^{-2}$\,s$^{-1}$   \\
$\chi^2$/d.o.f.                     & 359.00/118                &   32.98/37  (98.40)      &   \\ 
 nhp                                      & 3.77$\times10^{-26}$  &  0.658                         &   \\ 
\noalign{\smallskip}%
 \hline 
 \noalign{\smallskip} 
\multicolumn{4}{c}{{\tt tbabs * ztbabs * zpowerlw} (0.3--10\,keV)} \\     %
 \noalign{\smallskip} 
$N_{\rm H,z}$                          & $3.2_{-0.5}^{+0.6}$      & $2.6_{-2.6p}^{+5.3}$     &   $\times10^{21}$\,cm$^{-2}$             \\
$\Gamma_{\rm 0.3-10}$           & $1.75_{-0.09}^{+0.10}$ & $1.08_{-0.64}^{+0.71}$  &                \\
$F_{\rm 0.3-10}$                      &  8.71                         &  14.5                          &    $\times10^{-13}$ erg\,cm$^{-2}$\,s$^{-1}$   \\
$\chi^2$/d.o.f.                     & 142.83/117              &  33.96/36 (98.17)      &   \\ 
 nhp                                     &  0.0525                     &   0.566                       &   \\ 
\noalign{\smallskip}
 \hline 
 \noalign{\smallskip} 
\multicolumn{4}{c}{{\tt tbabs * zpcfabs * zpowerlw} (0.3--10\,keV)} \\ 
 \noalign{\smallskip} 
$N_{\rm H,z}$                          & $6.2\pm1.3$             &  --    &   $\times10^{21}$\,cm$^{-2}$             \\    
$f$                                        & $0.91_{-0.03}^{+0.02}$ &  --    &  \\
$\Gamma_{\rm 0.3-10}$           & $1.93\pm0.12$         &  --    &                \\
$F_{\rm 0.3-10}$                      &  8.54                          &   --   &    $\times10^{-13}$ erg\,cm$^{-2}$\,s$^{-1}$   \\
$\chi^2$/d.o.f.                     &  124.47/116              &  --  &   \\ 
 nhp                                      &  0.279                        &  --   &   \\ 
\noalign{\smallskip} 
 \hline 
 \noalign{\smallskip} 
\multicolumn{4}{c}{{\tt tbabs * absori * zpowerlw} (0.3--10\,keV)} \\ 
 \noalign{\smallskip} 
$N_{\rm H,z}$                          & $5.4_{-1.1}^{+1.2}$      &  --    &    $\times10^{21}$\,cm$^{-2}$            \\    
$\Gamma_{\rm 0.3-10}$           & $1.89_{-0.11}^{+0.12}$ &  --    &                \\
$\xi$                                    & $2.4_{-2.0}^{+2.5}$      &  --  &  $\times10^{-2}$ \\ 
$F_{\rm 0.3-10}$                      &  8.52                         &   --   &    $\times10^{-13}$ erg\,cm$^{-2}$\,s$^{-1}$   \\
$\chi^2$/d.o.f.                     & 128.51/116              &  --  &   \\ 
 nhp                                      &  0.201                       &  --   &   \\ 
\noalign{\smallskip}
 \hline 
 \noalign{\smallskip} 
\multicolumn{4}{c}{{\tt tbabs * zxipcf * zpowerlw} (0.3--10\,keV)} \\ 
 \noalign{\smallskip} 
$N_{\rm H,z}$                          & $6.2_{-1.8}^{+1.3}$       &  --    &   $\times10^{21}$\,cm$^{-2}$             \\    
$\log \xi$                           & $-0.63_{-1.07}^{+0.17}$ & --   & \\ 
$f$                                        & $0.99_{-0.04}^{+0.01p}$ &  --    &  \\
$\Gamma_{\rm 0.3-10}$           & $2.00\pm+0.17$ &  --    &                \\
$F_{\rm 0.3-10}$                      &  8.46                          &   --   &    $\times10^{-13}$ erg\,cm$^{-2}$\,s$^{-1}$   \\
$\chi^2$/d.o.f.                     &  127.88/115              &  --    &   \\ 
 nhp                                      &  0.194                        &  --    &   \\ 
\noalign{\smallskip}
 \hline
\noalign{\smallskip}
  \end{tabular}
  \end{center}
{\bf Notes.}  The fits are  for 
$z=0.16409$ and a fixed $N_{\rm H}^{\rm Gal}=1.4\times10^{20}$\,cm$^{-2}$ ({\sc tbabs}). 
Fluxes are corrected for the Galactic absorption. 
For the flare spectrum, Cash statistics was used and 
the goodness of fit (g.o.f.) was calculated with 10$^4$ simulations.  
Uncertainties are given at 90\,\% c.l.\ for one interesting parameter. 
 \end{table}

Given the lack of variations in the HR, 
we first considered two spectra, the one accumulated throughout the whole 2-yr campaign 
(`total', MJD range 58826--59423, $\sim$181\,ks exposure), 
and the `flare' spectrum,  close to the May 2020 radio flare (MJD  58994--58997,  $\sim$3.5\,ks).  

We fit the total spectrum  with {\sc XSPEC} \citep[v.\ 12.12.0,][]{Arnaud1996:xspec} 
with a simple power law model ({\sc zpowerlw}, fixed redshift $z=0.16409$)  
corrected for an equivalent hydrogen column corresponding to the Galactic value (fixed $N_{\rm H}^{\rm Gal}=1.4\times10^{20}$\,cm$^{-2}$, \citealt[][]{NH2016:HI4PI}) 
with the {\sc tbabs} \citep[][]{Wilms2000} model.  
A fit with a neutral absorber in the whole XRT band (0.3--10\,keV), 
yielding a hard photon index $\Gamma=1.13\pm0.04$, was clearly unacceptable 
(${\rm nhp}\sim10^{-26}$) due to the large negative residuals at energies below 2\,keV, 
suggesting further absorption is required. This is confirmed by the fact that a fit 
in the 2--10\,keV range yielded reasonable results and a much softer 
photon index, $\Gamma_{\rm 2-10\,keV}=1.84\pm0.15$  
(see Fig.~\ref{sdss1641:fig:tot_spec}b). 
The details of the fits are reported in Table~\ref{sdss1641:tab:swift_xrt_spec}. 
To model the excess absorption we added a second absorption component local to the source  ({\sc ztbabs}) 
obtaining  $N_{\rm H,z}=(3.2_{-0.5}^{+0.6})\times10^{21}$\,cm$^{-2}$ and $\Gamma =1.75_{-0.09}^{+0.10}$ (Fig.~\ref{sdss1641:fig:tot_spec}c). 
The trend in the residuals shown in Fig.~\ref{sdss1641:fig:tot_spec}c,  indicates that the model adopted for the extra neutral absorber 
may not be the most appropriate, so we considered a redshifted absorption edge ({\sc zedge}). 
The resulting fit, however, did not reach convergence, with an unphysical edge energy and an unconstrained optical depth, 
leading to even more prominent residuals below 1\,keV.    
Next, we considered a neutral absorber partially covering the central source, described by a {\sc zpcfabs}  component,
with the redshift of this absorber fixed at the same value as that of the source.  
This model yielded significant improvement ($\Delta \chi^2=-18$) over the  ({\sc ztbabs}) and, as shown in 
Fig.~\ref{sdss1641:fig:tot_spec}d, no large scale structure is left in the data/model ratio.  
The absorber is characterised by $N_{\rm H,z}=(6.2\pm1.3)\times10^{21}$\,cm$^{-2}$,  
a covering fraction $f=0.91_{-0.03}^{+0.02}$, 
while the underlying continuum has a photon index $\Gamma=1.93\pm0.12$. 

Finally, we considered the possibility that the absorber is ionised. 
When using an  {\sc absori} \citep[][]{Done1992,Magdziarz1995,Zdziarski1995} component,  
with fixed absorber temperature and Iron abundance, we obtain a significant improvement 
($\Delta \chi^2=-14$) over the  {\sc ztbabs} model, but a 
worse fit than the  {\sc zpcfabs} one.  
The presence of an ionised absorber partially covering the central source modelled {\sc zxipcf} is also 
not supported by the data, since 
the fit offers no improvement over the 
{\sc zpcfabs} model (see Table~\ref{sdss1641:tab:swift_xrt_spec}). 

We fit the flare spectrum, consisting of 42 photons, in the 0.3--10\,keV band, 
with the same models as those considered for the total spectrum (see Table~\ref{sdss1641:tab:swift_xrt_spec})
by adopting Cash \citep[][]{Cash79} statistics. 
The spectrum can be satisfactorily fit even without an added absorption component. 
Indeed,  for the {\sc tbabs * ztbabs * zpowerlw}, the {\sc tbabs * zpcfabs * zpowerlw}
and {\sc tbabs * zxipcf * zpowerlw} models the $N_{\rm H,z}$ is consistent with zero, 
for the {\sc tbabs * absori * zpowerlw} model the fit did not converge, with an ionisation parameter consistent with 0. 
We note, in particular, that if we fit the flare spectrum with the {\sc tbabs * ztbabs * zpowerlw} model 
with the photon index fixed to the mean value (1.75), we obtain  
$N_{\rm H,z}=(6.1^{+4.9}_{-3.0})\times10^{21}$\,cm$^{-2}$.   
This implies that we cannot exclude that extra absorption is indeed present. 

We conclude that the total spectrum of \corto{} can be best  described in the X-rays 
by an absorbed power law model modified by a partially covering neutral absorber  ({\sc tbabs * zpcfabs * zpowerlw})
with a covering fraction $f=0.91_{-0.03}^{+0.02}$ and underlying continuum with a photon index $\Gamma_{\rm total}=1.93\pm0.12$. 
Conversely, the flare spectrum can be satisfactorily fit even without an added absorption component by adopting 
a power-law model with a very hard photon index of $\Gamma_{\rm flare}=0.67^{+0.38}_{-0.39}$ 
modified just by the Galactic absorption.

Since the XRT data are spread over a 20-month baseline,  and guided by the examination of the XRT light curve,  
we also extracted spectra in several time intervals  as detailed in Table~\ref{sdss1641:tab:swift_xrt_spec_sel}:    
`1st yr':       all observations collected during the first year of the campaign  (MJD range 58826--59078,  on source exposure $\sim$113.3\,ks);   
`2nd yr':       all observations collected during the second year of the campaign  (MJD range 59245--59423,  on source exposure $\sim$67.8\,ks);   
`pre0':       all observations preceding the radio flare  of 2020-05-24  (MJD range 58826--58987,  on source exposure $\sim$61\,ks);   
`pre1':       close observations preceding the radio flare of 2020-05-24  (MJD  58938--58987, $\sim$22\,ks);          
`plateau': during an enhanced state of X-ray emission following the radio flare (MJD 59001--59029, $\sim$23\,ks); 
`post':      the remainder of the first year campaign  (MJD  59032--59078,  $\sim$24\,ks).                                           
Since no ToO observations were obtained after the April 2021 radio flare, 
recorded on 2021-04-07 and 2021-04-13, no XRT data are close enough ($\la 3$\,d) 
to be considered simultaneous; therefore  we did not perform any detailed spectroscopy. 
We fit all spectra with the same models as those adopted for the total spectrum described above, 
and the results are reported in full in Table~\ref{sdss1641:tab:swift_xrt_spec_full}. 
In the case of the  {\sc tbabs * zpcfabs * zpowerlw} model, we note that we 
obtained consistent covering factors in each time selected spectrum, therefore 
we also performed a fit with the covering factor fixed to the value derived from the total spectrum, 
$f=0.91$. 
In all cases a simple absorbed power-law model is not an adequate description, 
and an extra absorption component is required.

  \section{Discussion \label{sdss1641:discussion}}

We report on the first multi-wavelength \sw{} monitoring campaign performed on \src, 
a nearby ($z=0.16409$) NLS1 formerly known as radio-quiet 
which was however recently detected not only in the radio, 
but also in the $\gamma$-rays \citep[][]{Lahteenmaki2018}, 
a behaviour which hints at the presence of a relativistic jet. 
Our \sw{}  campaign, with a regular pace of one $\sim2$--3\,ks observation per week, lasted two years and  was performed with the primary goal of assessing the baseline variability properties (out of radio flare) of \corto{} 
and to obtain matching, multi-wavelength data strictly simultaneous 
with the 37\,GHz monitoring data being collected at \MH{} \citep[][]{Lahteenmaki2017}. 
Indeed, the campaign covered two distinct radio flaring episodes, one each year and 
for the first flare (2020-05-24 to  2020-05-26) we also obtained further ToO observations 
in order to have a denser sampling of the \sw{} light curves. 

             \subsection{Spectral variability and excess absorption}\label{sdss1641:discuss_spectrosc} 

From our time-selected X-ray spectroscopy (see Table~\ref{sdss1641:tab:swift_xrt_spec_full}) 
we find that the source is remarkably 
stable during the two years of monitoring, with the notable exception of the flare spectrum, 
extracted almost simultaneously with the radio flare in 2020. 
The mean spectrum, as represented by the total one in Table~\ref{sdss1641:tab:swift_xrt_spec},
is best described by an absorbed power law model ({\sc zpowerlw}, fixed redshift $z=0.16409$), 
where in addition to the Galactic absorption ({\sc tbabs}, with fixed $N_{\rm H}^{\rm Gal}=1.4\times10^{20}$\,cm$^{-2}$), 
an additional absorber is required. 
We model this component with a partially covering ($f=0.91_{-0.03}^{+0.02}$) neutral absorber,  
({\sc zpcfabs}, see Sect.~\ref{sdss1641:results_xrt}).  
The underlying continuum has a photon index $\Gamma_{\rm total}=1.93\pm0.12$
which places \corto{} in  the bulk of the radio loud $\gamma$-NLS1 population shown in figure~2 by 
\citet[][]{Foschini2015:fsrl_nls1}, whose median is $\Gamma=1.8$. 
Our conclusions stand when considering any of the time selections we performed, particularly 
those before and after the radio flare.  

On the contrary, the X-ray spectrum extracted around the radio flare does not require any 
absorption in excess of the Galactic one and is much harder, with a $\Gamma_{\rm flare} \sim 0.7\pm0.4$.
We cannot exclude extra absorption when fitting the flare spectrum with the {\sc tbabs * ztbabs * zpowerlw} model 
with the photon index fixed to the mean value (1.75). Indeed, 
we obtain $N_{\rm H,z}<11 \times10^{21}$\,cm$^{-2}$,  i.e., considerably higher than that of the mean spectrum.
Contrary to the averaged spectrum, in the flare case the fit shows a hard gamma also when restricted to the hard 
band, thus suggesting that the source is intrinsically harder.  
Although it is not unphysical to interpret the simultaneous increase of the X-ray flux  and absorption during the flare 
we consider more likely that we are seeing, instead,  
the presence of a harder spectral component distinct from the softer emission observed out of flare. 
We can naturally interpret these findings in terms of the jet emission emerging from a gap in the absorber. 
Consequently, this implies that, thanks to the observed correlated radio - X-ray variability, 
the radio emission during the flare is indeed linked to the presence of a jet in \corto,  
as opposed to out-of-flare radio emission which may be dominated by star formation activity 
\citep[][]{Berton2020:absorbed_jets}, as also testified by the WISE colours
$W1-W2=1$ and $W2-W3=4.3$ (as derived from NED) 
which place \corto{} outside of the AGN wedge 
(\citealt[][]{Mateos2012,Mateos2013}, see also figure 3 of \citealt{Foschini2015:fsrl_nls1} 
for a comparison of other radio loud NLS1 and radio loud $\gamma$-NLS1s).

We note that we also considered the possibility that the absorber is ionised, in line with the 
hypothesis put forth by \citet[][]{Berton2020:absorbed_jets} that the radio emission from the relativistic 
jet in \corto{} can be absorbed in the JVLA bands through free-free absorption due a screen of ionised gas associated 
with starburst activity (or shocks). 
And indeed an {\sc absori} component fits the X-ray data reasonably well, 
although not statistically as well as the neutral absorber model. 
More dedicated optical and X-ray spectral analysis will be presented in a companion paper 
(L\"ahteenm\"aki et al., in preparation), that should help clarify the nature of the absorber.

             \subsection{Spectral Energy Distributions}\label{sdss1641:discuss_seds} 

\begin{figure}[t] 
\vspace{-0.4truecm}

\hspace{-0.4truecm}
\includegraphics[angle=0,width=0.53\textwidth]{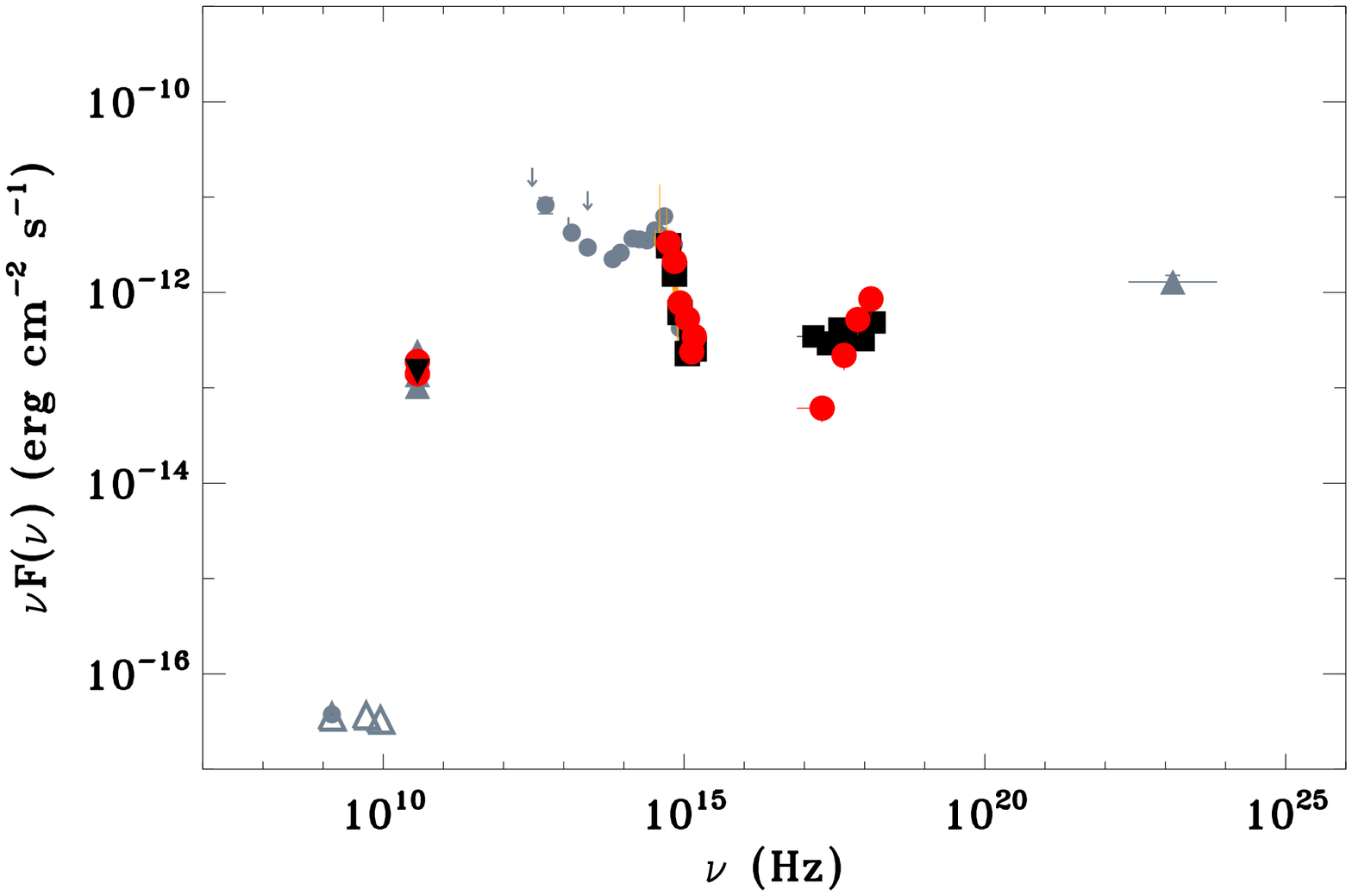}
  \caption{Spectral energy distribution of \src.
The black filled square points represent the (\sw) data obtained during the 2-yr \sw{} campaigns; 
the black filled triangle is  a representative \MH{} upper limit (0.40\,Jy) calculated as an average over 15\,d, while the red filled circles are those, strictly simultaneous, obtained during the flare of May 2020 (\sw{} and \MH). 
The SDSS spectrum is shown in orange. 
The grey points are drawn from the literature: 
JVLA 1.6, 5.2, and 9\,GHz data (empty triangles),  
\MH{} 37\,GHz data  (filled triangles),  
and \fermi{} (filled triangles), as well as 
FIRST, IRAS, WISE, USNO, 2MASS, SDSS, and WGA catalogues points collected from 
the ASI/SSDC SED Builder Tool.
 \label{sdss1641:fig:flare_SED} 
}
 \end{figure} 

In Fig.~\ref{sdss1641:fig:flare_SED} we show the radio to $\gamma$-ray spectral energy distribution (SED)
of \corto, 
that includes the data obtained during the 2-yr \sw{} campaigns,
covering the optical/UV + X-ray energy ranges and a representative upper limit from \MH{} (0.40\,Jy) calculated as an average over 15\,d, 
as well as  the data obtained during the May 2020 flare (\sw{} and \MH). 
We note that this flaring SED is the first strictly simultaneous broad-band SED (radio + optical/UV + X-ray) 
obtained for this source. 
The archival SDSS spectrum\footnote{\href{https://skyserver.sdss.org/dr17/}{https://skyserver.sdss.org/dr17/.}} is shown in orange.  
The grey points, instead, represent historical data drawn from the literature, and they include: 
JVLA \citep[1.6, 5.2, and 9\,GHz, empty triangles,][]{Berton2020:absorbed_jets}, 
\MH{}   \citep[37\,GHz, filled triangles,][]{Lahteenmaki2018},  
and \fermi{} \citep[filled triangles,][]{Lahteenmaki2018}. 
We also report data from FIRST, IRAS, WISE, USNO, 2MASS, SDSS,  and WGA catalogues as collected from 
the ASI/SSDC {\sc SED Builder Tool}\footnote{\href{https://tools.ssdc.asi.it/SED}
{https://tools.ssdc.asi.it/SED.}}  \citep[][]{Stratta2011:ASDC_SED_B} 
which include public catalogues and surveys.

The SED of \corto{} resembles those of other jetted sources, with a hint of a double-humped shape. 
However, we do not have a strictly simultaneous \fermi/LAT detection during the radio, optical-UV and X-ray flares, 
therefore it is difficult to constrain the whole SED. 
We note, however, that the SED of \corto{} resemble those of other \gray{} NLS1 galaxies, 
with a synchrotron peak below $\nu^{\rm peak}_{\rm syn} \approx 10^{13}$\,Hz, 
a host galaxy component peaking at a few $\times 10^{14}$\,Hz 
(as a comparison with a Sb galaxy template indicates) 
and the X-ray data which could be modelled with a synchrotron self-Compton component 
\citep[see e.g.,][]{Abdo2009:fit_sed_4nlsy,Foschini2015:fsrl_nls1}. 
While the optical-UV data during the May 2020 flare (red points) seem 
consistent with the average 2-yr \sw{} campaigns (black points), 
the 0.3--10\,keV data during the flare have a notably harder spectrum than 
the average one.

\begin{figure} 
\hspace{-0.5truecm}
\includegraphics[angle=-90,width=0.52\textwidth]{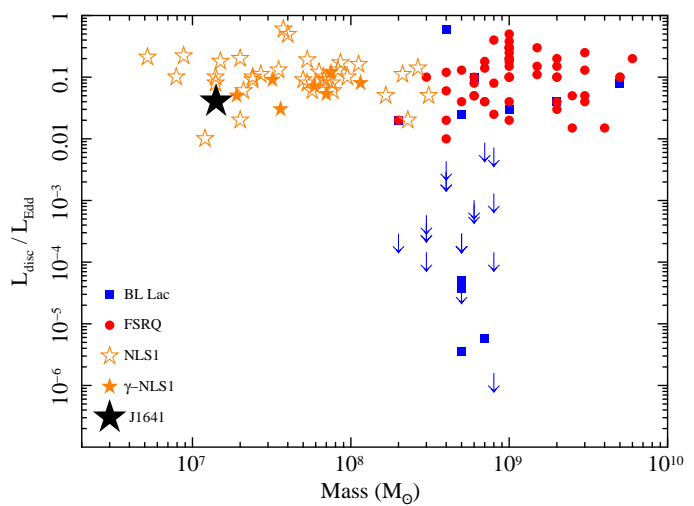}
  \caption{Accretion disc luminosity normalised to the Eddington luminosity as a function of the black-hole mass in solar masses. 
The black star indicates \src{}. Symbols and colors are the same as those in Figure~4 in ~\cite{Foschini2015:fsrl_nls1}.
\label{sdss1641:fig:f4LF15}
 }
 \end{figure} 

             \subsection{Energetics}\label{sdss1641:discuss_ene} 

In the following we shall compute quantities related to the physical processes occurring in \corto{}  
and  compare them with those of other \gray{} NLS1s. 
Following the procedure in \citet{Berton2021:PKS2004}, we first estimate the luminosity of the $\Hbeta$ emission line, 
$L_{\Hbeta}=2.51 \times 10^{41}$\,\ergsec, and
use it to calculate the distance of the  broad-line region (BLR) from \citet[][see Table~2]{Greene2010:RBLR} 
\begin{equation}
  \log \left[ \frac{R_{\rm BLR}}{\rm 10\,light\,day} \right]       =
   0.85 + 0.53 \log \left[ \frac{L_{\Hbeta}}{10^{43}\rm \ergsec} \right], 
\end{equation}
which implies $R_{\rm BLR} \simeq 2.6 \times 10^{16}$ \,{\rm cm}. 
The BLR luminosity 
is calculated according to   
\begin{equation}
  L_{\rm BLR}   =   21.2 \times L_{\Hbeta}\,(\ergsec)                                                 \simeq   5.3 \times 10^{42}\,\ergsec.
\end{equation}
The BLR radiation energy density is then derived from the BLR radius and luminosity, 
\begin{equation}
   u_{\rm BLR} = \frac{L_{\rm BLR}}{4\,\pi\,R^{2}_{\rm BLR}\,c}  {\rm (erg\,cm^{-3})}    \simeq 0.02\,\, {\rm erg\,cm^{-3}}.
\end{equation}
Given $R_{\rm BLR}$, we can estimate the accretion disc luminosity \citep[see e.g.,][]{Ghisellini2009}:
\begin{equation}
  L_{\rm disc} = 10^{45} \times \left(\frac{R_{\rm BLR}}{10^{17}\,{\rm cm}} \right)^2 (\ergsec)     \simeq 6.8 \times 10^{43}\,\, \ergsec,
\end{equation}
which can be normalised to the Eddington luminosity, 
\begin{equation}
    L_{\rm Edd} = 1.26\times10^{38}\left(\frac{M_{\rm BH}}{\Msun{}}\right) \, (\ergsec)                \simeq 1.8 \times 10^{45}\,\,\ergsec,
\end{equation}
assuming $M_{\rm BH} = 1.41 \times 10^{7}$\,\Msun{} as computed in~\cite{Jarvela2015}.
We obtain $L_{\rm disc}/L_{\rm Edd} \simeq 0.04$.

\setcounter{table}{3}		
  \begin{table} 	
 \begin{center} 	
 \caption{Overall properties of \src.	
 \label{sdss1641:tab:energetics}} 	
 \begin{tabular}{lcc } 
 \hline 
 \hline 
 \noalign{\smallskip} 
 Parameter & Value	 & Units	    \\
 \noalign{\smallskip} 
 \hline 
 \noalign{\smallskip}            
$M_{\rm BH}$$^{\mathrm{a}}$ & 1.41       & $\times10^{7}$\Msun  \\       
$L_{\rm H\beta}$         & 2.51       &  $\times10^{41}$\,erg\,s$^{-1}$  \\  
\noalign{\smallskip} 
 \hline 
 \noalign{\smallskip} 
$L_{\rm Edd}$	                &	1.8	  & $\times10^{45}$\,erg\,s$^{-1}$	\\
$L_{\rm disc}$	                &	6.8	  & $\times10^{43}$\,erg\,s$^{-1}$	\\
$L_{\rm disc}/L_{\rm Edd}$    &	0.04	  &	---			\\
$R_{\rm BLR}$	               &	2.6	  & $\times10^{16}$\,cm  			\\
$L_{\rm BLR}$	                &	5.3	  & $\times10^{42}$\,erg\,s$^{-1}$	\\
$u_{\rm BLR}$	                &	0.02	   & erg\,cm$^{-3}$	\\
$\alpha_{\rm 1.6-5.2\,GHz}$	&	1.04	   &	 ---			\\
$\alpha_{\rm 5.2-9.0\,GHz}$	&	1.24	    &	 ---			\\
$\alpha_{\rm 9.0-37\,GHz}$	&	-4.92  &	 ---			\\
$S_{\rm 15\,GHz}$	        &	4.33	    & mJy 			\\
$L_{\rm 15\,GHz}$	&	1.95	   & $\times10^{40}$\,erg\,s$^{-1}$	\\
$P^{\rm rad}_{\rm jet}$	&	1.65	   & $\times10^{42}$\,erg\,s$^{-1}$	\\
$P^{\rm kin}_{\rm jet}$  &	1.83	   & $\times10^{42}$\,erg\,s$^{-1}$	\\
$P^{\rm tot}_{\rm jet}$	 &	3.48	   & $\times10^{42}$\,erg\,s$^{-1}$	\\
  \noalign{\smallskip}
  \hline
  \end{tabular}
  \end{center}
  \begin{list}{}{} 
  \item[$^{\mathrm{a}}$] From \citet[][]{Jarvela2015}.
  \end{list} 
  \end{table}

Figure~\ref{sdss1641:fig:f4LF15} shows the position of \corto{} (black filled star) in the $L_{\rm disc}/L_{\rm Edd}$--$M_{\rm BH/\Msun}$ plane. 
Open (filled) orange stars mark the place of NLS1 (\gray-NLS1) objects, respectively. 
This plot is adapted from Figure~4 in~\cite{Foschini2015:fsrl_nls1} where red circles are FSRQs 
and the blue squares are BL~Lac objects (blue arrows indicate upper limits in the accretion luminosity). 
We note that \corto{} fits well in the region of the \gray{} NLS1 galaxies.

Another important quantity that characterises jetted sources is the jet power. 
\cite{Foschini2014:Pjet} established two relations between the radiative and 
kinetic jet power and the luminosity at 15\,GHz:
\begin{equation}
    \log(P^{\rm rad}_{\rm jet}) = (12 \pm 2) + (0.75 \pm 0.04)\,\log(L^{\rm core}_{\rm radio})\,\, (\ergsec),
\end{equation}
\begin{equation}
     \log(P^{\rm kin}_{\rm jet}) = (6 \pm 2) + (0.90 \pm 0.04)\,\log(L^{\rm core}_{\rm radio})\,\, (\ergsec).
\end{equation}
\cite{Jarvela2021:FrASS} provides simultaneous radio fluxes at 1.6, 5.2, 9.0\,GHz from the
 Karl G.\  Jansky Very Large Array (JVLA) and 37\,GHz data from the \MH{}  Radio Observatory 
(non-simultaneous with the JVLA). From these data we can estimate the radio spectral 
index\footnote{This value needs to be considered with caution as the data are not simultaneous.} 
$\alpha_{\rm 9.0-37\,GHz} \simeq -4.9$, obtaining $S_{\rm 15GHz} = 4.3$\, mJy, assuming
\begin{equation}
    \alpha_{1,2} = - \frac{\log(S_{1}/S_{2})}{\log(\nu_{1}/\nu_{2})},
\end{equation}
with $S_{9.0\rm GHz} = 0.35$\,mJy and $S_{37\rm GHz} = 370.0$\,mJy.
The $K$-corrected luminosity equation, 
\begin{equation}
    L = 4\,\pi\,D^{2}_{\rm lum}\,\nu\,S_{\nu}\,(1+z)^{(\alpha_{9.0-37\,\rm GHz}-1)},
\end{equation}
 yields $L_{\rm 15\,GHz} = 1.95 \times 10^{40}$\,\ergsec\, and allows us to estimate
$P^{\rm rad}_{\rm jet} = 1.65 \times 10^{42}$\,\ergsec, 
$P^{\rm kin}_{\rm jet} = 1.83 \times 10^{42}$\,\ergsec, 
and $P^{\rm tot}_{\rm jet} = 3.48 \times 10^{42}$\,\ergsec. 
We note that $\log(P^{\rm tot}_{\rm jet}) \simeq 42.54$\,\ergsec{} 
is similar to the lowest value in Table~3 of~\cite{Foschini2015:fsrl_nls1}, 
corresponding to J0706$+$3901, whose physical values 
\citep[see Table~2 in][]{Foschini2015:fsrl_nls1} are similar to those of \corto{}. 
In Table~\ref{sdss1641:tab:energetics} we summarise the results of our calculations.

  \section{Summary and conclusions \label{sdss1641:conclusions}}

In this paper we present  a  20-month {\em Swift} monitoring campaign performed, for the first time, 
on the newly recognized \citep[][]{Lahteenmaki2018} $\gamma$-NLS1 galaxy \src{} 
which is regularly being observed at the  \MH{}  Radio Observatory at the 37\,GHz frequency. 
Our investigation has indeed led us to the following findings and conclusions. 
\begin{itemize}
\item We observe minor, but significant variability in the light curves, with an F$_{\rm var}$, as calculated in
          the UVOT and XRT energy bands, that is smaller than those reported in the literature for 
          the best six well-known $\gamma$-NLS1s, whose light curves, however, are biased towards flaring states.
\item During the {\em Swift} campaign two radio flares were observed at \MH,  and for the first one (May 2020), 
          it was possible to obtain further  {\em Swift} observations closely matching the radio ones. 
\item From the analysis of the X-ray spectra preceding and following the radio flare we find that 
          a simple absorbed power-law model is not adequate and extra absorption is required.  Indeed, 
          the average spectrum of \corto{}  can be best  described
          by an absorbed power law model with a photon index $\Gamma=1.93\pm0.12$,
          modified by a partially covering neutral absorber (covering fraction $f=0.91_{-0.03}^{+0.02}$).
\item The X-ray spectrum closest to the radio flare, however,  does not require extra absorption and
          is much harder ($\Gamma_{\rm flare} \sim 0.7\pm0.4$). 
          This implies the emergence of a further, harder spectral component
          which we interpret as the jet emission peeking out of a gap in the absorber. 
\item For this flaring state, we present, also for the first time,  a strictly simultaneous SED covering 
          the radio, optical/UV, and X-ray energy bands afforded by the combination of 
          {\em Swift} and \MH{} data. 
          A comparison with the average SED (see Fig.~\ref{sdss1641:fig:flare_SED}) for this object, 
          highlights the harder X-ray spectrum during the radio flare. 
          No obvious variations are observed in the optical, however.  
\item Overall, the SED, although not well constrained at the high energies due to the lack of simultaneous 
          {\it Fermi}/LAT data, does show a resemblance to the SEDs of jetted sources, with hints at the presence of two humps. 
         The SED of \corto{} is reminiscent of other $\gamma$-NLS1 galaxies with a synchrotron
         peak below $10^{13}$\,Hz, a host galaxy component peaking at a few $10^{14}$\,Hz, 
         and the X-ray data which could be modelled with a synchrotron self-Compton component 
         \citep[][]{Abdo2009:fit_sed_4nlsy,Foschini2015:fsrl_nls1}.
\item Assuming that the radio emission is due to a jet, then we can calculate its power, 
          $\log(P^{\rm tot}_{\rm jet}) \simeq 42.54$\,\ergsec{}, which is one of the lowest measured when compared with the 
          \citet[][]{Foschini2015:fsrl_nls1} sample, and reminiscent of the $\gamma$-NLS1  J0706$+$3901. 
\end{itemize}

In conclusion, our observations 
show how a dedicated and well-paced monitoring campaign  
covering simultaneously the radio (at high frequencies) and the X-ray energy bands 
can allow us to interpret the source properties observed in different emission states.  
In particular, in the case of \src, 
we have been able to detect the `smoking-gun' of the so-called `absorbed jet', 
originally proposed by \citet[][]{Berton2020:absorbed_jets}. 
Indeed, the simultaneity and co-spatiality of the radio and X-ray emission show that the nucleus is responsible for the observed flaring of the source. 
Furthermore, if the origin of the flare were a putative extended emission, detected by  JVLA but unresolved by \MH, 
we would also observe a kpc-scale source with a variability timescale of days, which is not physically possibile.  
Currently, our knowledge of the absorbed jets is still limited, and only new observations in the X-rays will allow us to 
confirm if they all show signs of absorption.

  \begin{acknowledgements} 
      We thank the anonymous referee for comments that helped improve the paper.
      This work has been partially supported by the ASI-INAF program I/004/11/4. 
      We acknowledge financial contribution from the agreement ASI-INAF n.\ 2017-14-H.0.  
      This publication makes use of data obtained at Mets\"ahovi Radio Observatory, operated by Aalto University in Finland. 
       We acknowledge the use of public data from the {\em Swift} data archive.
      This research has made use of the NASA/IPAC Extragalactic Database (NED) which is operated by the Jet Propulsion Laboratory, 
      California Institute of Technology, under contract with the National Aeronautics and Space Administration.
      Part of this work is based on archival data, software or online services provided by the Space Science Data Center - ASI. 
      Happy 18th, \sw.
  \end{acknowledgements}


\begin{thebibliography}{}
\makeatletter
\relax
\def\mn@urlcharsother{\let\do\@makeother \do\$\do\&\do\#\do\^\do\_\do\%\do\~}
\def\mn@doi{\begingroup\mn@urlcharsother \@ifnextchar [ {\mn@doi@}
  {\mn@doi@[]}}
\def\mn@doi@[#1]#2{\def\@tempa{#1}\ifx\@tempa\@empty \href
  {http://dx.doi.org/#2} {doi:#2}\else \href {http://dx.doi.org/#2} {#1}\fi
  \endgroup}
\def\mn@eprint#1#2{\mn@eprint@#1:#2::\@nil}
\def\mn@eprint@arXiv#1{\href {http://arxiv.org/abs/#1} {{\tt arXiv:#1}}}
\def\mn@eprint@dblp#1{\href {http://dblp.uni-trier.de/rec/bibtex/#1.xml}
  {dblp:#1}}
\def\mn@eprint@#1:#2:#3:#4\@nil{\def\@tempa {#1}\def\@tempb {#2}\def\@tempc
  {#3}\ifx \@tempc \@empty \let \@tempc \@tempb \let \@tempb \@tempa \fi \ifx
  \@tempb \@empty \def\@tempb {arXiv}\fi \@ifundefined
  {mn@eprint@\@tempb}{\@tempb:\@tempc}{\expandafter \expandafter \csname
  mn@eprint@\@tempb\endcsname \expandafter{\@tempc}}}

\bibitem[\protect\citeauthoryear{{Abdo} et~al.,}{{Abdo}
  et~al.}{2009a}]{Abdo2009:J0948discov}
{Abdo} A.~A.,  et~al., 2009a, \mn@doi [\apj] {10.1088/0004-637X/699/2/976},
  \href {http://adsabs.harvard.edu/abs/2009ApJ...699..976A} {699, 976}

\bibitem[\protect\citeauthoryear{{Abdo} et~al.,}{{Abdo}
  et~al.}{2009b}]{Abdo2009:J0949mw}
{Abdo} A.~A.,  et~al., 2009b, \mn@doi [\apj] {10.1088/0004-637X/707/1/727},
  \href {http://adsabs.harvard.edu/abs/2009ApJ...707..727A} {707, 727}

\bibitem[\protect\citeauthoryear{{Abdo} et~al.,}{{Abdo}
  et~al.}{2009c}]{Abdo2009:fit_sed_4nlsy}
{Abdo} A.~A.,  et~al., 2009c, \mn@doi [\apjl] {10.1088/0004-637X/707/2/L142},
  \href {https://ui.adsabs.harvard.edu/abs/2009ApJ...707L.142A} {707, L142}

\bibitem[\protect\citeauthoryear{{Albareti} et~al.,}{{Albareti}
  et~al.}{2017}]{Albareti2017:SDSS13DR}
{Albareti} F.~D.,  et~al., 2017, \mn@doi [\apjs] {10.3847/1538-4365/aa8992},
  \href {https://ui.adsabs.harvard.edu/abs/2017ApJS..233...25A} {233, 25}

\bibitem[\protect\citeauthoryear{{Arnaud}}{{Arnaud}}{1996}]{Arnaud1996:xspec}
{Arnaud} K.~A.,  1996, in {Jacoby} G.~H.,  {Barnes} J.,  eds,  Astronomical
  Society of the Pacific Conference Series Vol. 101, Astronomical Data Analysis
  Software and Systems V. p.~17

\bibitem[\protect\citeauthoryear{{Berton} \& {J{\"a}rvel{\"a}}}{{Berton} \&
  {J{\"a}rvel{\"a}}}{2021}]{Berton2021:AN}
{Berton} M.,  {J{\"a}rvel{\"a}} E.,  2021, \mn@doi [Astronomische Nachrichten]
  {10.1002/asna.20210036}, \href
  {https://ui.adsabs.harvard.edu/abs/2021AN....342.1066B} {342, 1066}

\bibitem[\protect\citeauthoryear{{Berton} et~al.,}{{Berton}
  et~al.}{2016}]{Berton2016c}
{Berton} M.,  et~al., 2016, \mn@doi [\aap] {10.1051/0004-6361/201628171}, \href
  {https://ui.adsabs.harvard.edu/abs/2016A&A...591A..98B} {591, A98}

\bibitem[\protect\citeauthoryear{{Berton} et~al.,}{{Berton}
  et~al.}{2017}]{Berton2017:3C286}
{Berton} M.,  et~al., 2017, \mn@doi [Frontiers in Astronomy and Space Sciences]
  {10.3389/fspas.2017.00008}, \href
  {http://adsabs.harvard.edu/abs/2017FrASS...4....8B} {4, 8}

\bibitem[\protect\citeauthoryear{{Berton} et~al.,}{{Berton}
  et~al.}{2020}]{Berton2020:absorbed_jets}
{Berton} M.,  et~al., 2020, \mn@doi [\aap] {10.1051/0004-6361/202037793}, \href
  {https://ui.adsabs.harvard.edu/abs/2020A&A...636A..64B} {636, A64}

\bibitem[\protect\citeauthoryear{{Berton} et~al.,}{{Berton}
  et~al.}{2021}]{Berton2021:PKS2004}
{Berton} M.,  et~al., 2021, \mn@doi [\aap] {10.1051/0004-6361/202141409}, \href
  {https://ui.adsabs.harvard.edu/abs/2021A&A...654A.125B} {654, A125}

\bibitem[\protect\citeauthoryear{{Boller}, {Brandt}  \& {Fink}}{{Boller}
  et~al.}{1996}]{BollerBF1996:softX}
{Boller} T.,  {Brandt} W.~N.,   {Fink} H.,  1996, \aap, \href
  {http://adsabs.harvard.edu/abs/1996A%26A...305...53B} {305, 53}

\bibitem[\protect\citeauthoryear{{Brandt}, {Mathur}  \& {Elvis}}{{Brandt}
  et~al.}{1997}]{Brandt1997a}
{Brandt} W.~N.,  {Mathur} S.,   {Elvis} M.,  1997, \mn@doi [\mnras]
  {10.1093/mnras/285.3.L25}, \href
  {https://ui.adsabs.harvard.edu/abs/1997MNRAS.285L..25B} {285, L25}

\bibitem[\protect\citeauthoryear{{Burrows} et~al.,}{{Burrows}
  et~al.}{2005}]{Burrows2005:XRT}
{Burrows} D.~N.,  et~al., 2005, \mn@doi [\ssr] {10.1007/s11214-005-5097-2},
  \href {http://adsabs.harvard.edu/abs/2005SSRv..120..165B} {120, 165}

\bibitem[\protect\citeauthoryear{{Cash}}{{Cash}}{1979}]{Cash79}
{Cash} W.,  1979, \mn@doi [\apj] {10.1086/156922}, \href
  {http://adsabs.harvard.edu/abs/1979ApJ...228..939C} {228, 939}

\bibitem[\protect\citeauthoryear{{Cracco}, {Ciroi}, {Berton}, {Di Mille},
  {Foschini}, {La Mura}  \& {Rafanelli}}{{Cracco} et~al.}{2016}]{Cracco2016}
{Cracco} V.,  {Ciroi} S.,  {Berton} M.,  {Di Mille} F.,  {Foschini} L.,  {La
  Mura} G.,   {Rafanelli} P.,  2016, \mn@doi [\mnras] {10.1093/mnras/stw1689},
  \href {http://adsabs.harvard.edu/abs/2016MNRAS.462.1256C} {462, 1256}

\bibitem[\protect\citeauthoryear{{D'Ammando}}{{D'Ammando}}{2020}]{Dammando2020:swift}
{D'Ammando} F.,  2020, \mn@doi [\mnras] {10.1093/mnras/staa1580}, \href
  {https://ui.adsabs.harvard.edu/abs/2020MNRAS.496.2213D} {496, 2213}

\bibitem[\protect\citeauthoryear{{D'Ammando}, {Orienti}, {Finke}, {Larsson},
  {Giroletti}  \& {Raiteri}}{{D'Ammando} et~al.}{2016}]{Dammando2016:jets_nls1}
{D'Ammando} F.,  {Orienti} M.,  {Finke} J.,  {Larsson} J.,  {Giroletti} M.,
  {Raiteri} C.,  2016, \mn@doi [Galaxies] {10.3390/galaxies4030011}, \href
  {http://adsabs.harvard.edu/abs/2016Galax...4...11D} {4, 11}

\bibitem[\protect\citeauthoryear{{Done}, {Mulchaey}, {Mushotzky}  \&
  {Arnaud}}{{Done} et~al.}{1992}]{Done1992}
{Done} C.,  {Mulchaey} J.~S.,  {Mushotzky} R.~F.,   {Arnaud} K.~A.,  1992,
  \mn@doi [\apj] {10.1086/171649}, \href
  {https://ui.adsabs.harvard.edu/abs/1992ApJ...395..275D} {395, 275}

\bibitem[\protect\citeauthoryear{{Foschini}}{{Foschini}}{2012}]{Foschini2012:review}
{Foschini} L.,  2012, in Proceedings of Nuclei of Seyfert galaxies and QSOs -
  Central engine \& conditions of star formation (Seyfert 2012). 6-8 November,
  2012. Max-Planck-Insitut f{\"u}r Radioastronomie (MPIfR), Bonn, Germany.
  Online at <A href=``http://pos.sissa.it/cgi-bin/reader/conf.cgi?confid=169''>
  http://pos.sissa.it/cgi-bin/reader/conf.cgi?confid=169</A>, id.10. p.~10
  (\mn@eprint {arXiv} {1301.5785})

\bibitem[\protect\citeauthoryear{{Foschini}}{{Foschini}}{2014}]{Foschini2014:Pjet}
{Foschini} L.,  2014, in International Journal of Modern Physics Conference
  Series. p. 1460188 (\mn@eprint {arXiv} {1310.5822}),
  \mn@doi{10.1142/S2010194514601884}

\bibitem[\protect\citeauthoryear{{Foschini}, {Maraschi}, {Tavecchio},
  {Ghisellini}, {Gliozzi}  \& {Sambruna}}{{Foschini}
  et~al.}{2009}]{Foschini2009:Adv}
{Foschini} L.,  {Maraschi} L.,  {Tavecchio} F.,  {Ghisellini} G.,  {Gliozzi}
  M.,   {Sambruna} R.~M.,  2009, \mn@doi [Advances in Space Research]
  {10.1016/j.asr.2008.12.021}, \href
  {http://adsabs.harvard.edu/abs/2009AdSpR..43..889F} {43, 889}

\bibitem[\protect\citeauthoryear{{Foschini}, {Fermi/Lat Collaboration},
  {Ghisellini}, {Maraschi}, {Tavecchio}  \& {Angelakis}}{{Foschini}
  et~al.}{2010}]{Foschini2010:J0948}
{Foschini} L.,  {Fermi/Lat Collaboration} {Ghisellini} G.,  {Maraschi} L.,
  {Tavecchio} F.,   {Angelakis} E.,  2010, in {Maraschi} L.,  {Ghisellini} G.,
  {Della Ceca} R.,   {Tavecchio} F.,  eds,  Astronomical Society of the Pacific
  Conference Series Vol. 427, Accretion and Ejection in AGN: a Global View. pp
  243--248 (\mn@eprint {arXiv} {0908.3313})

\bibitem[\protect\citeauthoryear{{Foschini} et~al.,}{{Foschini}
  et~al.}{2015}]{Foschini2015:fsrl_nls1}
{Foschini} L.,  et~al., 2015, \mn@doi [\aap] {10.1051/0004-6361/201424972},
  \href {http://adsabs.harvard.edu/abs/2015A%26A...575A..13F} {575, A13}

\bibitem[\protect\citeauthoryear{{Foschini} et~al.,}{{Foschini}
  et~al.}{2021}]{Foschini2021:new_sample}
{Foschini} L.,  et~al., 2021, \mn@doi [Universe] {10.3390/universe7100372},
  \href {https://ui.adsabs.harvard.edu/abs/2021Univ....7..372F} {7, 372}

\bibitem[\protect\citeauthoryear{{Foschini} et~al.,}{{Foschini}
  et~al.}{2022}]{Foschini2022:new_sample2}
{Foschini} L.,  et~al., 2022, \mn@doi [Universe] {10.3390/universe8110587},
  \href {https://ui.adsabs.harvard.edu/abs/2022Univ....8..587F} {8, 587}

\bibitem[\protect\citeauthoryear{{Gehrels} et~al.,}{{Gehrels}
  et~al.}{2004}]{Gehrels2004}
{Gehrels} N.,  et~al., 2004, \mn@doi [\apj] {10.1086/422091}, \href
  {http://adsabs.harvard.edu/abs/2004ApJ...611.1005G} {611, 1005}

\bibitem[\protect\citeauthoryear{{Ghisellini} \& {Tavecchio}}{{Ghisellini} \&
  {Tavecchio}}{2009}]{Ghisellini2009}
{Ghisellini} G.,  {Tavecchio} F.,  2009, \mn@doi [\mnras]
  {10.1111/j.1365-2966.2009.15007.x}, \href
  {https://ui.adsabs.harvard.edu/abs/2009MNRAS.397..985G} {397, 985}

\bibitem[\protect\citeauthoryear{{Goodrich}}{{Goodrich}}{1989}]{Goodrich1989:nls1def}
{Goodrich} R.~W.,  1989, \mn@doi [\apj] {10.1086/167586}, \href
  {http://adsabs.harvard.edu/abs/1989ApJ...342..224G} {342, 224}

\bibitem[\protect\citeauthoryear{{Greene} et~al.,}{{Greene}
  et~al.}{2010}]{Greene2010:RBLR}
{Greene} J.~E.,  et~al., 2010, \mn@doi [\apj] {10.1088/0004-637X/723/1/409},
  \href {https://ui.adsabs.harvard.edu/abs/2010ApJ...723..409G} {723, 409}

\bibitem[\protect\citeauthoryear{{HI4PI Collaboration} et~al.,}{{HI4PI
  Collaboration} et~al.}{2016}]{NH2016:HI4PI}
{HI4PI Collaboration} et~al., 2016, \mn@doi [\aap]
  {10.1051/0004-6361/201629178}, \href
  {https://ui.adsabs.harvard.edu/abs/2016A&A...594A.116H} {594, A116}

\bibitem[\protect\citeauthoryear{{J{\"a}rvel{\"a}}, {L{\"a}hteenm{\"a}ki}  \&
  {Le{\'o}n-Tavares}}{{J{\"a}rvel{\"a}} et~al.}{2015}]{Jarvela2015}
{J{\"a}rvel{\"a}} E.,  {L{\"a}hteenm{\"a}ki} A.,   {Le{\'o}n-Tavares} J.,
  2015, \mn@doi [\aap] {10.1051/0004-6361/201424694}, \href
  {https://ui.adsabs.harvard.edu/abs/2015A&A...573A..76J} {573, A76}

\bibitem[\protect\citeauthoryear{{J{\"a}rvel{\"a}}, {L{\"a}hteenm{\"a}ki}  \&
  {Berton}}{{J{\"a}rvel{\"a}} et~al.}{2018}]{Jarvela2018}
{J{\"a}rvel{\"a}} E.,  {L{\"a}hteenm{\"a}ki} A.,   {Berton} M.,  2018, \mn@doi
  [\aap] {10.1051/0004-6361/201832876}, \href
  {https://ui.adsabs.harvard.edu/abs/2018A&A...619A..69J} {619, A69}

\bibitem[\protect\citeauthoryear{{J{\"a}rvel{\"a}}, {Berton}  \&
  {Crepaldi}}{{J{\"a}rvel{\"a}} et~al.}{2021}]{Jarvela2021:FrASS}
{J{\"a}rvel{\"a}} E.,  {Berton} M.,   {Crepaldi} L.,  2021, \mn@doi [Frontiers
  in Astronomy and Space Sciences] {10.3389/fspas.2021.735310}, \href
  {https://ui.adsabs.harvard.edu/abs/2021FrASS...8..147J} {8, 147}

\bibitem[\protect\citeauthoryear{{Komatsu} et~al.,}{{Komatsu}
  et~al.}{2011}]{Komatsu2011_cosmology}
{Komatsu} E.,  et~al., 2011, \mn@doi [\apjs] {10.1088/0067-0049/192/2/18},
  \href {https://ui.adsabs.harvard.edu/abs/2011ApJS..192...18K} {192, 18}

\bibitem[\protect\citeauthoryear{{Komossa}, {Voges}, {Xu}, {Mathur}, {Adorf},
  {Lemson}, {Duschl}  \& {Grupe}}{{Komossa} et~al.}{2006}]{Komossa2006:rlnl1q}
{Komossa} S.,  {Voges} W.,  {Xu} D.,  {Mathur} S.,  {Adorf} H.-M.,  {Lemson}
  G.,  {Duschl} W.~J.,   {Grupe} D.,  2006, \mn@doi [\aj] {10.1086/505043},
  \href {http://adsabs.harvard.edu/abs/2006AJ....132..531K} {132, 531}

\bibitem[\protect\citeauthoryear{{L{\"a}hteenm{\"a}ki}
  et~al.,}{{L{\"a}hteenm{\"a}ki} et~al.}{2017}]{Lahteenmaki2017}
{L{\"a}hteenm{\"a}ki} A.,  et~al., 2017, \mn@doi [\aap]
  {10.1051/0004-6361/201630257}, \href
  {http://adsabs.harvard.edu/abs/2017A%26A...603A.100L} {603, A100}

\bibitem[\protect\citeauthoryear{{L{\"a}hteenm{\"a}ki}, {J{\"a}rvel{\"a}},
  {Ramakrishnan}, {Tornikoski}, {Tammi}, {Vera}  \&
  {Chamani}}{{L{\"a}hteenm{\"a}ki} et~al.}{2018}]{Lahteenmaki2018}
{L{\"a}hteenm{\"a}ki} A.,  {J{\"a}rvel{\"a}} E.,  {Ramakrishnan} V.,
  {Tornikoski} M.,  {Tammi} J.,  {Vera} R.~J.~C.,   {Chamani} W.,  2018,
  \mn@doi [\aap] {10.1051/0004-6361/201833378}, \href
  {http://adsabs.harvard.edu/abs/2018A%26A...614L...1L} {614, L1}

\bibitem[\protect\citeauthoryear{{Leighly}}{{Leighly}}{1999}]{Leighly1999b}
{Leighly} K.~M.,  1999, \mn@doi [\apjs] {10.1086/313287}, \href
  {https://ui.adsabs.harvard.edu/abs/1999ApJS..125..317L} {125, 317}

\bibitem[\protect\citeauthoryear{{Magdziarz} \& {Zdziarski}}{{Magdziarz} \&
  {Zdziarski}}{1995}]{Magdziarz1995}
{Magdziarz} P.,  {Zdziarski} A.~A.,  1995, \mn@doi [\mnras]
  {10.1093/mnras/273.3.837}, \href
  {https://ui.adsabs.harvard.edu/abs/1995MNRAS.273..837M} {273, 837}

\bibitem[\protect\citeauthoryear{{Marziani} et~al.,}{{Marziani}
  et~al.}{2018}]{Marziani2018:FrASS}
{Marziani} P.,  et~al., 2018, \mn@doi [Frontiers in Astronomy and Space
  Sciences] {10.3389/fspas.2018.00006}, \href
  {https://ui.adsabs.harvard.edu/abs/2018FrASS...5....6M} {5, 6}

\bibitem[\protect\citeauthoryear{{Mateos} et~al.,}{{Mateos}
  et~al.}{2012}]{Mateos2012}
{Mateos} S.,  et~al., 2012, \mn@doi [\mnras]
  {10.1111/j.1365-2966.2012.21843.x}, \href
  {https://ui.adsabs.harvard.edu/abs/2012MNRAS.426.3271M} {426, 3271}

\bibitem[\protect\citeauthoryear{{Mateos}, {Alonso-Herrero}, {Carrera},
  {Blain}, {Severgnini}, {Caccianiga}  \& {Ruiz}}{{Mateos}
  et~al.}{2013}]{Mateos2013}
{Mateos} S.,  {Alonso-Herrero} A.,  {Carrera} F.~J.,  {Blain} A.,  {Severgnini}
  P.,  {Caccianiga} A.,   {Ruiz} A.,  2013, \mn@doi [\mnras]
  {10.1093/mnras/stt953}, \href
  {https://ui.adsabs.harvard.edu/abs/2013MNRAS.434..941M} {434, 941}

\bibitem[\protect\citeauthoryear{{Mathur}}{{Mathur}}{2000}]{Mathur2000}
{Mathur} S.,  2000, \mn@doi [\mnras] {10.1046/j.1365-8711.2000.03530.x}, \href
  {https://ui.adsabs.harvard.edu/abs/2000MNRAS.314L..17M} {314, L17}

\bibitem[\protect\citeauthoryear{{Olgu{\'\i}n-Iglesias}, {Kotilainen}  \&
  {Chavushyan}}{{Olgu{\'\i}n-Iglesias}
  et~al.}{2020}]{Olguin-Iglesias2020_HostClass}
{Olgu{\'\i}n-Iglesias} A.,  {Kotilainen} J.,   {Chavushyan} V.,  2020, \mn@doi
  [\mnras] {10.1093/mnras/stz3549}, \href
  {https://ui.adsabs.harvard.edu/abs/2020MNRAS.492.1450O} {492, 1450}

\bibitem[\protect\citeauthoryear{{Oshlack}, {Webster}  \& {Whiting}}{{Oshlack}
  et~al.}{2001}]{Oshlack2001:pks2004-447}
{Oshlack} A.~Y.~K.~N.,  {Webster} R.~L.,   {Whiting} M.~T.,  2001, \mn@doi
  [\apj] {10.1086/322299}, \href
  {http://adsabs.harvard.edu/abs/2001ApJ...558..578O} {558, 578}

\bibitem[\protect\citeauthoryear{{Osterbrock} \& {Pogge}}{{Osterbrock} \&
  {Pogge}}{1985}]{OsterbrockP1985:nls1def}
{Osterbrock} D.~E.,  {Pogge} R.~W.,  1985, \mn@doi [\apj] {10.1086/163513},
  \href {http://adsabs.harvard.edu/abs/1985ApJ...297..166O} {297, 166}

\bibitem[\protect\citeauthoryear{{Paliya} et~al.,}{{Paliya}
  et~al.}{2020}]{Paliya2020}
{Paliya} V.~S.,  et~al., 2020, \mn@doi [\apj] {10.3847/1538-4357/ab754f}, \href
  {https://ui.adsabs.harvard.edu/abs/2020ApJ...892..133P} {892, 133}

\bibitem[\protect\citeauthoryear{{Peterson} et~al.,}{{Peterson}
  et~al.}{2004}]{Peterson2004}
{Peterson} B.~M.,  et~al., 2004, \mn@doi [\apj] {10.1086/423269}, \href
  {http://adsabs.harvard.edu/abs/2004ApJ...613..682P} {613, 682}

\bibitem[\protect\citeauthoryear{{Romano}, {Vercellone}, {Foschini},
  {Tavecchio}, {Landoni}  \& {Kn{\"o}dlseder}}{{Romano}
  et~al.}{2018}]{Romano2018:nls1_cta}
{Romano} P.,  {Vercellone} S.,  {Foschini} L.,  {Tavecchio} F.,  {Landoni} M.,
   {Kn{\"o}dlseder} J.,  2018, \mn@doi [\mnras] {10.1093/mnras/sty2484}, \href
  {http://adsabs.harvard.edu/abs/2018MNRAS.481.5046R} {481, 5046}

\bibitem[\protect\citeauthoryear{{Roming} et~al.,}{{Roming}
  et~al.}{2005}]{Roming2005:UVOT}
{Roming} P.~W.~A.,  et~al., 2005, \mn@doi [\ssr] {10.1007/s11214-005-5095-4},
  \href {http://adsabs.harvard.edu/abs/2005SSRv..120...95R} {120, 95}

\bibitem[\protect\citeauthoryear{{Shao}, {Gu}, {Chen}, {Yang}, {Yao}, {Yuan}
  \& {Shen}}{{Shao} et~al.}{2022}]{Shao2022}
{Shao} X.,  {Gu} M.,  {Chen} Y.,  {Yang} H.,  {Yao} S.,  {Yuan} W.,   {Shen}
  Z.,  2022, arXiv e-prints, \href
  {https://ui.adsabs.harvard.edu/abs/2022arXiv221116705S} {p. arXiv:2211.16705}

\bibitem[\protect\citeauthoryear{{Stratta}, {Capalbi}, {Giommi}, {Primavera},
  {Cutini}, {Gasparrini}  \& {on behalf of the ASDC team}}{{Stratta}
  et~al.}{2011}]{Stratta2011:ASDC_SED_B}
{Stratta} G.,  {Capalbi} M.,  {Giommi} P.,  {Primavera} R.,  {Cutini} S.,
  {Gasparrini} D.,   {on behalf of the ASDC team} 2011, preprint, \href
  {http://cdsads.u-strasbg.fr/abs/2011arXiv1103.0749S} {} (\mn@eprint {arXiv}
  {1103.0749})

\bibitem[\protect\citeauthoryear{{Sulentic} \& {Marziani}}{{Sulentic} \&
  {Marziani}}{2015}]{SulenticMarziani2015:FrASS}
{Sulentic} J.,  {Marziani} P.,  2015, \mn@doi [Frontiers in Astronomy and Space
  Sciences] {10.3389/fspas.2015.00006}, \href
  {https://ui.adsabs.harvard.edu/abs/2015FrASS...2....6S} {2, 6}

\bibitem[\protect\citeauthoryear{{Sulentic}, {Marziani}, {Zamanov}, {Bachev},
  {Calvani}  \& {Dultzin-Hacyan}}{{Sulentic} et~al.}{2002}]{Sulentic2002}
{Sulentic} J.~W.,  {Marziani} P.,  {Zamanov} R.,  {Bachev} R.,  {Calvani} M.,
  {Dultzin-Hacyan} D.,  2002, \mn@doi [\apjl] {10.1086/339594}, \href
  {https://ui.adsabs.harvard.edu/abs/2002ApJ...566L..71S} {566, L71}

\bibitem[\protect\citeauthoryear{{Ter\"asranta} et~al.,}{{Ter\"asranta}
  et~al.}{1998}]{Teraesranta98}
{Ter\"asranta} H.,  et~al., 1998, \mn@doi [\aaps] {10.1051/aas:1998297}, \href
  {http://adsabs.harvard.edu/abs/1998A%26AS..132..305T} {132, 305}

\bibitem[\protect\citeauthoryear{{Varglund}, {J{\"a}rvel{\"a}},
  {L{\"a}hteenm{\"a}ki}, {Berton}, {Ciroi}  \& {Congiu}}{{Varglund}
  et~al.}{2022}]{Varglund2022}
{Varglund} I.,  {J{\"a}rvel{\"a}} E.,  {L{\"a}hteenm{\"a}ki} A.,  {Berton} M.,
  {Ciroi} S.,   {Congiu} E.,  2022, \mn@doi [\aap]
  {10.1051/0004-6361/202244465}, \href
  {https://ui.adsabs.harvard.edu/abs/2022A&A...668A..91V} {668, A91}

\bibitem[\protect\citeauthoryear{{Vaughan}, {Edelson}, {Warwick}  \&
  {Uttley}}{{Vaughan} et~al.}{2003}]{Vaughan2003}
{Vaughan} S.,  {Edelson} R.,  {Warwick} R.~S.,   {Uttley} P.,  2003, \mn@doi
  [\mnras] {10.1046/j.1365-2966.2003.07042.x}, \href
  {https://ui.adsabs.harvard.edu/abs/2003MNRAS.345.1271V} {345, 1271}

\bibitem[\protect\citeauthoryear{{Viswanath}, {Stalin}, {Rakshit}, {Kurian},
  {Ujjwal}, {Gudennavar}  \& {Kartha}}{{Viswanath}
  et~al.}{2019}]{Viswanath2019:NLS1masses}
{Viswanath} G.,  {Stalin} C.~S.,  {Rakshit} S.,  {Kurian} K.~S.,  {Ujjwal} K.,
  {Gudennavar} S.~B.,   {Kartha} S.~S.,  2019, \mn@doi [\apjl]
  {10.3847/2041-8213/ab365e}, \href
  {https://ui.adsabs.harvard.edu/abs/2019ApJ...881L..24V} {881, L24}

\bibitem[\protect\citeauthoryear{{Wilms}, {Allen}  \& {McCray}}{{Wilms}
  et~al.}{2000}]{Wilms2000}
{Wilms} J.,  {Allen} A.,   {McCray} R.,  2000, \mn@doi [\apj] {10.1086/317016},
  \href {http://adsabs.harvard.edu/abs/2000ApJ...542..914W} {542, 914}

\bibitem[\protect\citeauthoryear{{Yuan}, {Zhou}, {Komossa}, {Dong}, {Wang},
  {Lu}  \& {Bai}}{{Yuan} et~al.}{2008}]{Yuan2008}
{Yuan} W.,  {Zhou} H.~Y.,  {Komossa} S.,  {Dong} X.~B.,  {Wang} T.~G.,  {Lu}
  H.~L.,   {Bai} J.~M.,  2008, \mn@doi [\apj] {10.1086/591046}, \href
  {http://adsabs.harvard.edu/abs/2008ApJ...685..801Y} {685, 801}

\bibitem[\protect\citeauthoryear{{Zdziarski}, {Johnson}, {Done}, {Smith}  \&
  {McNaron-Brown}}{{Zdziarski} et~al.}{1995}]{Zdziarski1995}
{Zdziarski} A.~A.,  {Johnson} W.~N.,  {Done} C.,  {Smith} D.,   {McNaron-Brown}
  K.,  1995, \mn@doi [\apjl] {10.1086/187716}, \href
  {https://ui.adsabs.harvard.edu/abs/1995ApJ...438L..63Z} {438, L63}

\bibitem[\protect\citeauthoryear{{Zhou}, {Wang}, {Dong}, {Zhou}  \&
  {Li}}{{Zhou} et~al.}{2003}]{Zhou2003:0948}
{Zhou} H.-Y.,  {Wang} T.-G.,  {Dong} X.-B.,  {Zhou} Y.-Y.,   {Li} C.,  2003,
  \mn@doi [\apj] {10.1086/345523}, \href
  {http://adsabs.harvard.edu/abs/2003ApJ...584..147Z} {584, 147}

\bibitem[\protect\citeauthoryear{{Zhou}, {Wang}, {Yuan}, {Lu}, {Dong}, {Wang}
  \& {Lu}}{{Zhou} et~al.}{2006}]{Zhou2006:SDSS_NLS1s}
{Zhou} H.,  {Wang} T.,  {Yuan} W.,  {Lu} H.,  {Dong} X.,  {Wang} J.,   {Lu} Y.,
   2006, \mn@doi [\apjs] {10.1086/504869}, \href
  {https://ui.adsabs.harvard.edu/abs/2006ApJS..166..128Z} {166, 128}

\makeatother
\end{thebibliography}


\begin{appendix}
\section{Supplementary tables and figures}

\setcounter{table}{0} 
\begin{table*} 	 
 \begin{center} 	
 \caption{{\it Swift}/XRT observation log of \src. For the first year monitoring we report observing sequence, date (MJD of the middle of the observation), start and end times (UTC), and XRT exposure time.} 	
 \label{sdss1641:tab:swift_xrt_log} 	
 \begin{tabular}{lllll}
 \hline 
 \hline 
 \noalign{\smallskip} 
 Sequence   & Date  & Start time  (UT)  & End time   (UT) & Exposure  \\ 
           &   MJD (mid)       & (yyyy-mm-dd hh:mm:ss)  & (yyyy-mm-dd hh:mm:ss)  &(s)       \\
 \noalign{\smallskip} 
 \hline 
 \noalign{\smallskip}  
00011395003	&	58826.13783	&	2019-12-09 01:31:03	&	2019-12-09 05:05:53	&	2502	\\
00011395004	&	58833.07074	&	2019-12-16 00:39:51	&	2019-12-16 02:43:52	&	2959	\\
00011395005	&	58840.58190	&	2019-12-23 13:04:59	&	2019-12-23 14:50:53	&	2688	\\
00011395006	&	58847.64448	&	2019-12-30 09:09:11	&	2019-12-30 21:46:54	&	2913	\\
00011395007	&	58854.17679	&	2020-01-06 00:07:15	&	2020-01-06 08:21:53	&	2166	\\
00011395008	&	58861.57894	&	2020-01-13 12:09:26	&	2020-01-13 15:37:54	&	3009	\\
00011395009	&	58868.89399	&	2020-01-20 19:39:46	&	2020-01-20 23:14:54	&	2620	\\
00011395010	&	58875.75798	&	2020-01-27 17:18:31	&	2020-01-27 19:04:26	&	1198	\\
00011395011	&	58878.85233	&	2020-01-30 20:13:49	&	2020-01-30 20:40:52	&	1622	\\
00011395013	&	58889.60558	&	2020-02-10 08:05:12	&	2020-02-10 20:58:52	&	3249	\\
00011395014	&	58896.07517	&	2020-02-17 00:51:35	&	2020-02-17 02:44:53	&	2332	\\
00011395015	&	58903.10809	&	2020-02-24 01:45:18	&	2020-02-24 03:25:59	&	1036	\\
00011395017	&	58910.44720	&	2020-03-02 02:37:02	&	2020-03-02 18:50:54	&	2740	\\
00011395018	&	58917.85224	&	2020-03-09 17:54:33	&	2020-03-09 22:59:53	&	2766	\\
00011395019	&	58924.30035	&	2020-03-16 03:14:06	&	2020-03-16 11:10:53	&	3084	\\
00011395020	&	58931.50555	&	2020-03-23 10:21:05	&	2020-03-23 13:54:53	&	2465	\\
00011395021	&	58938.50374	&	2020-03-30 11:08:53	&	2020-03-30 13:01:52	&	2670	\\
00011395022	&	58945.20770	&	2020-04-06 04:02:16	&	2020-04-06 05:55:54	&	2735	\\
00011395023	&	58952.08435	&	2020-04-13 00:23:01	&	2020-04-13 03:39:54	&	2786	\\
00011395024	&	58958.45545	&	2020-04-19 09:04:47	&	2020-04-19 12:46:53	&	2833	\\
00011395025	&	58966.92002	&	2020-04-27 21:03:45	&	2020-04-27 23:05:53	&	3162	\\
00011395026	&	58973.76286	&	2020-05-04 17:17:08	&	2020-05-04 19:19:52	&	2771	\\
00011395027	&	58980.43370	&	2020-05-11 08:40:08	&	2020-05-11 12:08:55	&	2630	\\
00011395028	&	58987.48241	&	2020-05-18 00:14:28	&	2020-05-18 22:54:52	&	2397	\\
00011395029	&	58994.64205	&	2020-05-25 15:10:11	&	2020-05-25 15:38:54	&	1723	\\
00011395030$^{\mathrm{a}}$	&	58996.16110	&	2020-05-27 03:49:06	&	2020-05-27 03:54:52	&	346	\\
00011395032$^{\mathrm{a}}$	&	58997.30146	&	2020-05-28 00:31:22	&	2020-05-28 13:56:53	&	1484	\\
00011395033$^{\mathrm{a}}$	&	58999.38572	&	2020-05-30 02:03:05	&	2020-05-30 16:27:46	&	1504	\\
00011395034	&	59001.38639	&	2020-06-01 01:50:55	&	2020-06-01 16:41:52	&	3365	\\
00011395035$^{\mathrm{a}}$	&	59006.16589	&	2020-06-05 19:03:53	&	2020-06-06 12:53:52	&	2344	\\
00011395036$^{\mathrm{a}}$	&	59007.69577	&	2020-06-07 09:21:56	&	2020-06-08 00:01:53	&	2079	\\
00011395037	&	59008.34756	&	2020-06-08 07:22:05	&	2020-06-08 09:18:52	&	2949	\\
00011395038$^{\mathrm{a}}$	&	59009.90842	&	2020-06-09 20:15:22	&	2020-06-09 23:20:53	&	2129	\\
00011395039$^{\mathrm{a}}$	&	59011.17429	&	2020-06-11 04:06:03	&	2020-06-11 04:15:52	&	589	\\
00011395040	&	59015.69333	&	2020-06-15 11:38:54	&	2020-06-15 21:37:53	&	2608	\\
00011395041	&	59022.49924	&	2020-06-22 09:21:53	&	2020-06-22 14:35:54	&	2705	\\
00011395042$^{\mathrm{a}}$	&	59025.58544	&	2020-06-25 09:12:09	&	2020-06-25 18:53:53	&	1229	\\
00011395043	&	59029.60794	&	2020-06-29 11:59:58	&	2020-06-29 17:10:53	&	2630	\\
00011395044$^{\mathrm{a}}$	&	59032.92782	&	2020-07-02 21:16:13	&	2020-07-02 23:15:53	&	2655	\\
00011395045	&	59036.21290	&	2020-07-06 00:15:16	&	2020-07-06 09:57:52	&	2680	\\
00011395046$^{\mathrm{a}}$	&	59039.52393	&	2020-07-09 09:26:49	&	2020-07-09 15:42:06	&	1319	\\
00011395047	&	59043.34032	&	2020-07-13 07:10:15	&	2020-07-13 09:09:52	&	3016	\\
00011395048$^{\mathrm{a}}$	&	59046.52746	&	2020-07-16 05:16:12	&	2020-07-16 20:02:53	&	1647	\\
00011395049	&	59050.35234	&	2020-07-20 06:37:52	&	2020-07-20 10:16:52	&	2344	\\
00011395050	&	59057.34706	&	2020-07-27 07:19:39	&	2020-07-27 09:19:53	&	3056	\\
00011395051	&	59064.65346	&	2020-08-03 11:24:03	&	2020-08-03 19:57:54	&	2873	\\
00011395052	&	59071.58888	&	2020-08-10 12:27:25	&	2020-08-10 15:48:33	&	1176	\\
00011395053	&	59074.64927	&	2020-08-13 15:29:00	&	2020-08-13 15:40:52	&	712	\\
00011395054	&	59078.43118	&	2020-08-17 03:45:55	&	2020-08-17 16:55:52	&	2766	\\
  \noalign{\smallskip}
  \hline
  \end{tabular}
  \end{center}
  \begin{list}{}{} 
  \item[$^{\mathrm{a}}$] Data obtained through additional ToOs in response to the \MH{} detection on 2020-05-24.
  \end{list} 
  \end{table*}

\setcounter{table}{1} 
\begin{table*} 	 
 \begin{center} 	
   \caption{{\it Swift}/XRT observation log of \src. For the second year monitoring we report observing sequence, date (MJD of the middle of the observation), start and end times (UTC), and XRT exposure time.} 	
 \label{sdss1641:tab:swift_xrt_log2} 	
 \begin{tabular}{lllll}
 \hline 
 \hline 
 \noalign{\smallskip} 
 Sequence   & Date  & Start time  (UT)  & End time   (UT) & Exposure  \\ 
           &   MJD        & (yyyy-mm-dd hh:mm:ss)  & (yyyy-mm-dd hh:mm:ss)  &(s)       \\
 \noalign{\smallskip} 
 \hline 
 \noalign{\smallskip} 
00011395055	&	59245.65652	&	2021-01-31 14:00:52	&	2021-01-31 17:29:53	&	2698	\\
00011395056	&	59252.36437	&	2021-02-07 02:13:29	&	2021-02-07 15:15:52	&	2848	\\
00011395057	&	59259.69355	&	2021-02-14 15:41:32	&	2021-02-14 17:35:52	&	2743	\\
00011395058	&	59266.49648	&	2021-02-21 00:36:59	&	2021-02-21 23:12:53	&	2776	\\
00011395059	&	59273.73557	&	2021-02-28 15:49:32	&	2021-02-28 19:28:53	&	2778	\\
00011395060	&	59280.93881	&	2021-03-07 21:33:52	&	2021-03-07 23:29:54	&	2846	\\
00011395061	&	59287.71487	&	2021-03-14 14:32:54	&	2021-03-14 19:45:54	&	2209	\\
00011395062	&	59294.52133	&	2021-03-21 09:20:25	&	2021-03-21 15:40:52	&	2944	\\
00011395063	&	59301.51512	&	2021-03-28 11:26:00	&	2021-03-28 13:18:52	&	1745	\\
00011395064	&	59308.35169	&	2021-04-04 07:29:58	&	2021-04-04 09:22:53	&	2630	\\
00011395065	&	59315.91731	&	2021-04-11 21:04:58	&	2021-04-11 22:56:52	&	2658	\\
00011395066	&	59322.55583	&	2021-04-18 12:21:55	&	2021-04-18 14:18:51	&	2994	\\
00011395067	&	59329.76132	&	2021-04-25 16:32:42	&	2021-04-25 19:59:53	&	2357	\\
00011395068	&	59336.60402	&	2021-05-02 09:31:43	&	2021-05-02 19:27:51	&	1705	\\

00011395069	&	59339.52500	&	2021-05-05 12:33:08	&	2021-05-05 12:38:51	&	344	\\
00011395070	&	59343.31562	&	2021-05-09 01:05:06	&	2021-05-09 14:03:52	&	2964	\\
00011395071	&	59350.73554	&	2021-05-16 15:54:28	&	2021-05-16 19:23:52	&	2389	\\
00011395072	&	59357.94372	&	2021-05-23 21:47:01	&	2021-05-23 23:30:53	&	2101	\\
00011395073	&	59360.49007	&	2021-05-26 11:37:31	&	2021-05-26 11:53:51	&	980	\\
00011395074	&	59363.58868	&	2021-05-29 00:29:31	&	2021-05-30 03:45:52	&	1971	\\
00011395075	&	59371.64900	&	2021-06-06 13:47:13	&	2021-06-06 17:21:53	&	2209	\\
00011395076	&	59379.58760	&	2021-06-14 13:14:24	&	2021-06-14 14:57:53	&	2131	\\
00011395077	&	59385.45068	&	2021-06-20 07:30:58	&	2021-06-20 14:06:59	&	1522	\\
00011395078	&	59388.67917	&	2021-06-23 15:25:06	&	2021-06-23 17:10:53	&	1439	\\
00011395079	&	59392.72476	&	2021-06-27 16:29:23	&	2021-06-27 18:17:54	&	2733	\\
00011395080	&	59399.82883	&	2021-07-04 17:12:08	&	2021-07-04 22:34:53	&	2866	\\
00011395081	&	59406.49739	&	2021-07-11 10:11:35	&	2021-07-11 13:40:52	&	2788	\\
00011395082	&	59413.33648	&	2021-07-18 06:15:08	&	2021-07-18 09:53:54	&	2455	\\
00011395083	&	59420.43584	&	2021-07-25 10:14:19	&	2021-07-25 10:40:54	&	1595	\\
00011395084	&	59423.62380	&	2021-07-28 14:46:40	&	2021-07-28 15:09:52	&	1392	\\

 \noalign{\smallskip}
  \hline
  \end{tabular}
  \end{center}
  \end{table*}

\setcounter{table}{2}
\begin{table}
 \tabcolsep 2pt 
 \begin{center}
 \caption{Time-selected {\it Swift}/XRT spectroscopy: time ranges. }
\scriptsize
\small
 \label{sdss1641:tab:swift_xrt_spec_sel}
 \begin{tabular}{lcccccc c }
 \hline
 \hline
 \noalign{\smallskip}
 Spectr.   &   ObsID          & Start time (UT)           & End time (UT)             & Expo.       &    \\ 
  Name    &  range            &                                  &                                   &  (ks)          &      \\
  \noalign{\smallskip}
\hline
\noalign{\smallskip}
total                              & 003--084 &  2019-12-09 01:31:04 & 2021-07-28 15:09:52 &  181.07    &    \\ 
 1st year                        & 003--054 &  2019-12-09 01:31:04 & 2020-08-17 16:55:53 &  113.3       &    \\ 
 2nd year                       & 055--084 &  2021-01-31 14:00:52 & 2021-07-28 15:09:52 &   67.8        &    \\ 
\noalign{\smallskip}
   pre0                           & 003--028 &  2019-12-09 01:31:04 & 2020-05-18 22:54:52 &  61.334     &    \\ 
   pre1                           & 021--028 &  2020-03-30 11:08:54 & 2020-05-18 22:54:52 &  21.984     &    \\ 
  flare                            & 029--032 &  2020-05-25 15:10:12 & 2020-05-28 13:56:49 & 3.553        &    \\ 
  plateau                       & 034--043 &  2020-06-01 01:50:56 & 2020-06-29 17:10:54 &  22.626     &    \\ 
   post                           & 044--054 &  2020-07-02 21:16:14 & 2020-08-17 16:55:53 &  24.246    &    \\ 
 \noalign{\smallskip}
 \hline
\noalign{\smallskip}
\end{tabular}
  \end{center}
 \end{table}

\vfill
\newpage

\setcounter{table}{3}
\begin{table}[t]
 \tabcolsep 2pt 
 \begin{center}
 \caption{Time-selected {\it Swift}/XRT spectroscopy: results.  
The spectral fits are obtained by assuming a source redhift of $z=0.16409$ and 
a Galactic absorption modelled with a fixed {\sc tbabs} component with $N_{\rm H}^{\rm Gal}=1.4\times10^{20}$\,cm$^{-2}$. 
Absorption columns $N_{\rm H,z}$ are in units of $\times10^{21}$\,cm$^{-2}$, 
the ionization parameter $\xi$ in units of $\times10^{-2}$, while the 
observed fuxes are in units of $\times10^{-13}$\,erg\,cm$^{-2}$\,s$^{-1}$. 
All quoted uncertainties are given at 90\,\% confidence level for one interesting parameter. For the fits of the flare spectrum, Cash statistics was used and 
the goodness of fit (g.o.f.) was calculated with 10$^4$ simulations.}
\scriptsize
\small
 \label{sdss1641:tab:swift_xrt_spec_full}
 \begin{tabular}{ccccccc c }
 \hline 
 \hline 
 \noalign{\smallskip}
                                 &       & \multicolumn{3}{c}{  {\bf tbabs * zpowerlw} (2--10\,keV)}  &  \\
                    & &$\Gamma_{\rm 2-10}$ & $F_{\rm 2-10}$                   & $\chi^2$/d.o.f. &     \\  
 \noalign{\smallskip}
 \noalign{\smallskip}
 total                             &   &   $1.84_{-0.15}^{+0.15}$ & 6.23 &   42.40/48               &      \\
 1st year                        &   &   $1.75_{-0.21}^{+0.22}$ &  6.55 &   28.15/31              &     \\
 2nd year                       &   & $1.77_{-0. 23}^{+0. 23}$ & 5.84 &  36.08/17               &    \\
\noalign{\smallskip}
   pre0                             & & $1.79_{-0.33}^{+0.34}$ & 6.14 &  17.41/15                  &     \\
   pre1                             & & $1.70_{-0.73}^{+0.79}$ & 5.88 &  5.02/3                      &     \\
  flare                             & & $0.94_{-1.35}^{+1.35}$  & 14.1 &  \multicolumn{2}{l}{12.72/20 (70.48)}     &     \\
  plateau                         & & $1.93_{-0.61}^{+0.62}$  & 6.5   &  4.01/4                     &    \\
   post                             & & $1.61_{-0.67}^{+0.68}$ &  7.15 &  3.49/4                     &   \\
\noalign{\smallskip}
 \hline 
\noalign{\smallskip}

                    & & \multicolumn{3}{c}{{\bf tbabs * zpowerlw} (0.3--10\,keV)}  & & \\
 \noalign{\smallskip}
                     & & $\Gamma_{\rm 0.3-10}$ & $F_{\rm 0.3-10}$                   & $\chi^2$/d.o.f.              &             \\ 
   \noalign{\smallskip}
 \noalign{\smallskip}
 flare             & & $0.67_{-0.39}^{+0.38} $  & 17.2                                   &   32.98/37  (98.40)       &            \\   
 \noalign{\smallskip}
 \hline 
\noalign{\smallskip}

                                      & \multicolumn{5}{c}{{\bf tbabs * ztbabs * zpowerlw} (0.3--10\,keV)} \\
 \noalign{\smallskip}
                                      & $N_{\rm H,z}$              & $\Gamma_{\rm 0.3-10}$       & $F_{\rm 0.3-10}$                 & $\chi^2$/d.o.f. \\ 
  \noalign{\smallskip}
  \noalign{\smallskip}
 total                              & $3.2_{-0.5}^{+0.6}$  & $1.75_{-0.09}^{+0.10}$    &   8.64    &    142.83/117      \\
 1st year                         & $2.8_{-0.5}^{+0.6}$  & $1.72_{-0.11}^{+0.12}$    &    9.00     &    87.05/79      \\
 2nd year                         & $3.8_{-1.0}^{+1.2}$  & $1.78_{-0.16}^{+0.17}$    &    8.00    &     70.11/44       \\
\noalign{\smallskip}
   pre0                            & $2.6_{-0.7}^{+0.9}$  &  $1.71_{-0.16}^{+0.17}$   &   8.57     &    40.80/41         \\
   pre1                            & $2.7_{-1.2}^{+1.9}$   &  $1.66_{-0.27}^{+0.31}$   &   8.11     &    13.40/12          \\ 
  flare                             & $2.6_{-2.6p}^{+5.3}$   &  $1.08_{-0.64}^{+0.71}$  &   14.4      & \multicolumn{2}{l}{33.96/36 (98.17)}  \\ 
  plateau                        &  $2.2_{-0.9}^{+1.2}$  &  $1.76_{-0.24}^{+0.26}$   &    9.95     & 20.99/17            \\ 
   post                            & $3.2_{-1.4}^{+2.0}$ & $1.65_{-0.29}^{+0.33}$    &    9.32     &  17.62/16            \\ 
\noalign{\smallskip}
 \hline 
\noalign{\smallskip}

                                       & \multicolumn{5}{c}{{\bf tbabs * absori * zpowerlw} (0.3--10\,keV)} \\
 \noalign{\smallskip}

                                       & $N_{\rm H,z}$              & $\Gamma_{\rm 0.3-10}$     &  $\xi$        & $F_{\rm 0.3-10}$                 & $\chi^2$/d.o.f. \\ 
 \noalign{\smallskip}
  \noalign{\smallskip}
total                             & $5.4_{-1.1}^{+1.2}$   &  $1.89_{-0.11}^{+0.12}$    &$2.4_{-2.0}^{+2.5}$ &  8.45    &  128.51/116     \\
 1st year                       & $4.3_{-1.4}^{+1.3}$  & $1.83_{-0.15}^{+0.14}$     & $1.1_{-1.0}^{+2.9}$ &   8.80   &     83.11/78      \\
 2nd year                      & $6.6_{-2.1}^{+2.3}$  & $1.94_{-0.19}^{+0.20}$     & $4.3_{-3.0}^{+9.4}$ &   7.81    &      62.08/43       \\  
  \noalign{\smallskip}
   pre0                           & $4.2_{-2.1}^{+1.8}$   &  $1.84_{-0.23}^{+0.20}$    &$1.4_{-1.4p}^{+4.0}$ &  8.35    &   39.50/40    \\ 
   pre1                           & $5.8_{-4.0}^{+5.5}$   &  $1.89_{-0.42}^{+0.48}$    &$5.0_{-5.0p}^{+5.7}$ &  7.73   &   11.98/11    \\ 
  plateau                       & $4.1_{-2.5}^{+3.5}$   &  $1.92_{-0.32}^{+0.36}$    &$1.8_{-1.8p}^{+10.5}$ & 9.60     &  19.52/16     \\1 
   post                          & $6.4_{-3.3}^{+3.9}$   &  $1.88_{-0.36}^{+0.39}$    &$6.4_{-6.4}^{+40.8}$ &   8.78   &   14.43/15    \\
\noalign{\smallskip}
 \hline 
\noalign{\smallskip}

  & \multicolumn{5}{c}{{\bf tbabs * zpcfabs * zpowerlw} (0.3--10\,keV)} \\
\noalign{\smallskip}
 
                             & $N_{\rm H,z}$              & $f$                  &   $\Gamma_{\rm 0.3-10}$     &  $F_{\rm 0.3-10}$                 & $\chi^2$/d.o.f. \\

 \noalign{\smallskip}
 \noalign{\smallskip}
total                  & $6.2_{-1.3}^{+1.3}$&$0.91_{-0.03}^{+0.02}$ &$1.93_{-0.12}^{+0.12}$ &8.45 & 124.47/116\\
 1st year              & $4.7_{-1.5}^{+1.5}$&$0.91_{-0.05}^{+0.06}$ & $1.84_{-0.15}^{+0.15}$& 8.82& 82.25/78\\    
 2nd year              & $7.6_{-2.3}^{+2.5}$&$0.92_{-0.04}^{+0.03}$ & $2.00_{-0.20}^{+0.21}$& 7.81& 59.48/43 \\ 
  \noalign{\smallskip}
   pre0                  & $4.1_{-1.6}^{+2.3}$&$0.92_{-0.07}^{+0.01p}$ & $1.81_{-0.20}^{+0.23}$& 8.42& 39.41/40\\ 
   pre1                   &$6.2_{-4.2}^{+5.7}$ &$0.90_{-0.14}^{+0.01p}$  &$1.90_{-0.22}^{+0.51}$ &7.76 &12.26/11 \\
   plateau               & $5.3_{-1.8}^{+4.2}$&$0.88_{-0.12}^{+0.01p}$ & $1.98_{-0.36}^{+0.41}$&9.56 &19.19/16 \\
  post                    & $6.8_{-3.5}^{+4.1}$&$0.91_{-0.11}^{+0.07}$ & $1.91_{-0.38}^{+0.42}$&8.77 &14.42/15 \\
\noalign{\smallskip}
 \hline
\noalign{\smallskip}
  
 \noalign{\smallskip}
 \noalign{\smallskip}
 1st year              & $4.6_{-1.2}^{+1.5}$&$0.91$ & $1.84\pm0.15$& 8.82& 82.26/79\\   
 2nd year              & $7.6_{-2.2}^{+2.4}$&$0.91$ & $1.99_{-0.18}^{+0.15}$& 7.84& 59.50/44 \\ 
  \noalign{\smallskip}
   pre0                  & $4.2_{-1.5}^{+2.2}$&$0.91$ & $1.81_{-0.20}^{+0.22}$& 8.42& 39.42/41\\  
   pre1                   &$6.2_{-4.1}^{+5.4}$ &$0.91$  &$1.93_{-0.48}^{+0.35}$ &7.65 &12.38/12 \\
   plateau               & $4.7_{-2.9}^{+5.1}$&$0.91$ & $2.00_{-0.41}^{+0.38}$&9.41 &19.69/17 \\
   post                    & $6.9_{-3.6}^{+3.8}$&$0.91$ & $1.93_{-0.39}^{+0.27}$&8.69 &14.45/16 \\

\noalign{\smallskip}
 \hline
\noalign{\smallskip}

  \end{tabular}
  \end{center}
   \end{table}

\end{appendix}
\end{document}